\documentclass[aps,floatfix,twocolumn,superscriptaddress,amsmath,amssymb,showpacs,10pt]{revtex4-2}
\usepackage{graphicx}
\usepackage{epstopdf}
\usepackage{epsfig}
\usepackage{epsf}
\usepackage{url}
\usepackage[USenglish]{babel}
\usepackage{hyperref}
\def\bcen{\begin{center}}
\def\ecen{\end{center}}
\allowdisplaybreaks
\usepackage{verbatim}
\usepackage{natbib}
\usepackage{amsmath, nccmath}
\usepackage[export]{adjustbox}
\usepackage{dcolumn}
\usepackage{bm}
\usepackage{bbm}
\usepackage{lipsum}
\usepackage{color}
\usepackage{newtxtext}
\usepackage{newtxmath}
\usepackage{sidecap}
\usepackage[section]{placeins}

\newcommand{\ii}{\mathrm{i}}

\begin{document}
	
	\title{Nonequilibrium susceptibilities near ordering instabilities: transient short-range order, thermalization, and finite-size effects in correlated lattice systems}

	\author{Nathan Perruchoud}
	\affiliation{Department of Physics, University of Fribourg, 1700 Fribourg, Switzerland}
	\author{Maksymilian \'Sroda}
	\affiliation{Department of Physics, University of Fribourg, 1700 Fribourg, Switzerland}
	\author{Philipp Werner}
	\affiliation{Department of Physics, University of Fribourg, 1700 Fribourg, Switzerland}

\begin{abstract}
    We study the real-time dynamics of susceptibilities and nonlocal correlations close to ordering instabilities in two half-filled models on the square lattice: the Hubbard model within the fluctuation-exchange approximation and an electron--phonon model with acoustic phonons within the Migdal approximation.
    By compressing the nonequilibrium Green's functions in time as quantics tensor trains, we reach lattices of up to $256\times256$ sites. Starting from a cold noninteracting state, the Hubbard interaction or the electron--phonon coupling is ramped up to a value for which the thermalized system is close to an antiferromagnetic or charge-density-wave instability. In both models, the corresponding susceptibility at $(\pi,\pi)$ is transiently enhanced, and the correlation length grows by several lattice spacings before decreasing to the thermalized value. Effective temperatures extracted from the distribution functions show a slow relaxation toward the thermal state. Our results indicate that two ingredients are sufficient to transiently enhance short-range order
    in correlated lattice systems: 
    proximity to an ordering instability and a ramp which is faster than the heating of the electrons. The effect depends strongly on the lattice size, which makes large lattices, and hence compressed representations of the Green's functions, essential for numerical investigations.
\end{abstract}

\maketitle
	
\section{Introduction}

The nonequilibrium dynamics of interacting many-body systems exhibits rich phenomena, including prethermalization \cite{Berges2004,Moeckel2008,Sroda2025}, where local observables look thermalized, while nonlocal ones are far from thermal, relaxation bottlenecks \cite{Sensarma2010,Eckstein2011,Sentef2013,Murakami2015,Avigo2016,Sroda2026}, where populations get stuck in energy or momentum space due to a lack of relaxation channels, and various types of dynamical phase transitions \cite{Eckstein2009,Schiro2011} marked by qualitative changes in the relaxation dynamics, or nonanalyticities in return probabilities \cite{Heyl2014,Canovi2014}. In systems with broken symmetries, the proximity to nonthermal fixed points \cite{Berges2008,Tsuji2013} and the presence of energy gaps \cite{Werner2012} can transiently stabilize ordered states, while suitable excitations may even induce novel types of order \cite{Li2018,Kaneko2019,Li2020,Werner2021,Ray2023,Ray2024}. Many of these effects have been studied in model calculations with local self-energy approximations \cite{Freericks2006,Aoki2014}, which lack a feedback from nonlocal correlations on the evolution of the system. Such a feedback is, however, important in the vicinity of ordering instabilities, where collective effects and the presence of short-range order may affect the dynamics. For example, Ref.~\cite{Bauer2015} demonstrated the transient enhancement of antiferromagnetic (AFM) correlations after interaction ramps in the two-dimensional Hubbard model. In moderately correlated systems, this physics and related phenomena can be captured by weak-coupling diagrammatic calculations \cite{Bickers1989,Hedin1965,Migdal1958} with suitable momentum-dependent self-energies. These schemes are well established, but their nonequilibrium  implementation on the Kadanoff--Baym contour and application to large lattice systems pose significant numerical challenges \cite{Schuler2019}. The recent combination of nonequilibrium Green's function and quantics tensor-train compression techniques \cite{Shinaoka2023,Murray2024,Sroda2025,Sroda2026}, however, opens the possibility to simulate the nonequilibrium dynamics of susceptibilities and nonlocal correlations on lattices of unprecedented size. 

Here, we take advantage of these recent developments to study the transient evolution of correlated electron and electron--phonon systems in the vicinity of an ordering instability. Specifically, we consider the two-dimensional Hubbard model within the self-consistent fluctuation-exchange approximation \cite{Bickers1989}, and simulate the evolution of spin correlations in interaction ramps close to the AFM instability. As a second example, we solve the Fr\"ohlich Hamiltonian with acoustic phonons within the self-consistent Migdal approximation \cite{Migdal1958} and compute the evolution of charge correlations in phonon-coupling ramps in proximity to the charge-density-wave (CDW) instability. In both models, we find a transient enhancement of the corresponding susceptibility at $(\pi,\pi)$ and of the correlation length. We also find a strong lattice-size dependence of the nonequilibrium susceptibilities. Reliable simulations therefore require large lattices, which can only be treated with compressed Green's functions. Finally, we investigate the transient states and nonequilibrium dynamics by measuring effective temperatures for different correlation functions.     

This paper is organized as follows. Section~\ref{sec:model_method} introduces the two models, the fluctuation-exchange and Migdal approximations, and their numerical implementation with quantics tensor trains. In Sec.~\ref{results}, we present the results. After specifying the setup (Sec.~\ref{sec:setup}), we discuss the transient enhancement of the antiferromagnetic and charge-density-wave responses (Sec.~\ref{sec:transient}), the finite-size effects (Sec.~\ref{sec:finite_size}), the correlation length (Sec.~\ref{sec:xi}), and the effective temperatures (Sec.~\ref{sec:temperature}). Section~\ref{conclusions} summarizes our findings and discusses possible extensions of the method.

\section{Models and Method}\label{sec:model_method}

\subsection{Models}
We consider two interacting many-body systems on the square lattice at half filling. The first is a correlated electron system described by the Hubbard model~\cite{Hubbard1963}
\begin{equation}
H(t) = \sum_{\mathbf{k}\sigma} \varepsilon_{\mathbf{k}}\, c^\dagger_{\mathbf{k}\sigma} c_{\mathbf{k}\sigma}
+ U(t) \sum_{\mathbf{i}} \Bigl(n_{\mathbf{i}\uparrow}-\tfrac12\Bigr)\Bigl(n_{\mathbf{i}\downarrow}-\tfrac12\Bigr),
\end{equation}
where $\varepsilon_{\mathbf{k}} = -2t_h(\cos k_x + \cos k_y)$ is the dispersion, $t_h$ is the hopping parameter, $U(t)$ is the time-dependent interaction, and $c^\dagger_{\mathbf{k}\sigma}$ ($c_{\mathbf{k}\sigma}$) creates (annihilates) an electron with momentum $\mathbf{k}$ and spin $\sigma$. The operator $n_{\mathbf{i}\sigma}=c^\dagger_{\mathbf{i}\sigma}c_{\mathbf{i}\sigma}$ measures the spin density on site $\mathbf{i}$.

The second model describes electron--phonon interactions of Fr\"ohlich type~\cite{Frohlich1950},
\begin{equation}
\begin{split}
H(t) ={}& \sum_{\mathbf{k}\sigma} \varepsilon_{\mathbf{k}}\, c^\dagger_{\mathbf{k}\sigma} c_{\mathbf{k}\sigma}
+ \sum_{\mathbf{q}} \Omega_{\mathbf{q}}\, b^\dagger_{\mathbf{q}} b_{\mathbf{q}}\\
&+ \frac{1}{\sqrt{N_k}} \sum_{\mathbf{k}\mathbf{q}\sigma} g_{\mathbf{q}}(t)\, c^\dagger_{\mathbf{k}+\mathbf{q}\sigma} c_{\mathbf{k}\sigma} \bigl(b^\dagger_{-\mathbf{q}} + b_{\mathbf{q}}\bigr),
\end{split}
\label{eq:H_eph}
\end{equation}
where $b^\dagger_{\mathbf{q}}$ creates a phonon with momentum $\mathbf{q}$ and energy $\Omega_{\mathbf{q}}$, $N_k$ is the total number of momentum points, and $g_{\mathbf{q}}(t)$ is the electron--phonon coupling. We consider acoustic phonons,
\begin{equation}
\Omega_{\mathbf{q}} = \frac{\Omega_0}{\sqrt{2}}\sqrt{\sin^2\tfrac{q_x}{2}+\sin^2\tfrac{q_y}{2}},
\qquad
g_{\mathbf{q}}(t) = \sqrt{\Omega_{\mathbf{q}}}\, g(t),
\label{eq:phonon_dispersion}
\end{equation}
where $g(t)$ is the time-dependent coupling strength, $\Omega_0$ is the phonon energy at $\mathbf{q}=(\pi,\pi)$, and $g_{\mathbf{q}}$ vanishes as $\sqrt{q}$ for $q\to0$, as for the deformation-potential coupling to acoustic phonons~\cite{BardeenShockley1950,Migdal1958}. In both models, the system starts in a noninteracting thermal state, and the interaction $U(t)$ or the coupling $g(t)$ is switched on by a linear ramp.

\subsection{Nonequilibrium Green's functions}\label{sec:negf}
We use the nonequilibrium Green's function (NEGF) formalism~\cite{Aoki2014,Schlunzen2020,Stefanucci2025}, in which the thermal initial state and the dynamics of the system are described by two-time correlation functions defined on the three-branch Kadanoff--Baym contour $\mathcal{C}$, which runs along $0 \to t_{\max} \to 0 \to -\ii\beta$, with $\beta$ the inverse temperature of the initial state.
The fermionic Green's function for momentum $\mathbf{k}$ is
\begin{equation}
G_{\mathbf{k}}(t,t') = -\ii \bigl\langle \mathcal{T}_{\mathcal C}\, c_{\mathbf{k}\sigma}(t)\, c^\dagger_{\mathbf{k}\sigma}(t') \bigr\rangle,
\end{equation}
where $\mathcal{T}_{\mathcal C}$ denotes the contour-ordering operator. We consider paramagnetic states, so that $G_{\mathbf{k}}$ does not depend on $\sigma$. The Green's function obeys the Dyson equation
\begin{equation}
G_{\mathbf{k}} = G^0_{\mathbf{k}} + G^0_{\mathbf{k}} * \Sigma_{\mathbf{k}} * G_{\mathbf{k}},
\label{eq_dyson}
\end{equation}
where $*$ denotes a convolution along the contour, $G^0_{\mathbf{k}}$ is the noninteracting Green's function, and the self-energy $\Sigma_{\mathbf{k}}$ contains all interaction effects.

In this work, we mainly use two real-time components of these functions. For a contour-ordered two-time function $X(t,t') = -\ii\langle\mathcal{T}_{\mathcal C}\,A(t)\,B(t')\rangle$, the lesser component is $X^<(t,t') = \pm\ii\langle B(t')\,A(t)\rangle$ and the retarded component is $X^R(t,t') = -\ii\,\theta(t-t')\langle[A(t),B(t')]_\pm\rangle$, where the upper (lower) sign applies to fermionic (bosonic) functions, $[\cdot,\cdot]_+$ is the anticommutator, $[\cdot,\cdot]_-$ the commutator, and $\theta(t)$ the Heaviside step function \cite{Aoki2014}. The lesser component measures the correlations between times $t$ and $t'$, whereas the retarded component describes the response at time $t$ to a perturbation at time $t'$.

\subsection{Fluctuation-exchange approximation}\label{sec:flex}
For the Hubbard model, we use the weak-coupling fluctuation-exchange (FLEX) approximation~\cite{Bickers1989,Bickers1989b}, which has been previously applied to nonequilibrium systems in Refs.~\cite{Sayyad2019,Dasari2019}. Here, the self-energy in real space is defined as
\begin{equation}
\begin{split}
\Sigma_{\mathbf{ij}}(t,t') ={}& \ii\,U(t)U(t')\,G_{\mathbf{ij}}(t,t')\\
&\times\Bigl[\tfrac12\chi^c_{\mathbf{ij}} + \tfrac32\chi^s_{\mathbf{ij}} - \chi_{\mathbf{ij}}\Bigr](t,t'),
\end{split}
\label{eq:sigma_flex}
\end{equation}
where
\begin{equation}
\chi_{\mathbf{ij}}(t,t') = -\ii\,G_{\mathbf{ij}}(t,t')\,G_{\mathbf{ji}}(t',t)
\end{equation}
is the polarization bubble. The weight $\tfrac32$ counts the three equivalent spin channels and $\tfrac12$ the charge channel. Subtracting $\chi$ avoids counting the second-order diagram twice.
The charge and spin susceptibilities $\chi^c$ and $\chi^s$ are obtained from the bosonic Dyson equations
\begin{equation}
\chi^c_{\mathbf{q}} = \chi_{\mathbf{q}} + \chi_{\mathbf{q}} * U * \chi^c_{\mathbf{q}},
\qquad
\chi^s_{\mathbf{q}} = \chi_{\mathbf{q}} - \chi_{\mathbf{q}} * U * \chi^s_{\mathbf{q}},
\label{eq:bosonic_dyson}
\end{equation}
where $U$ stands for $U(t)\,\delta_{\mathcal C}(t,t')$. These RPA-type resummations, which do not include all vertex corrections~\cite{Bickers1989b,Monthoux1994},
provide approximations for the charge and spin correlation functions
\begin{equation}
\begin{aligned}
\chi^c_{\mathbf{ij}}(t,t') &\simeq -\tfrac{\ii}{2}\Bigl[\bigl\langle \mathcal T_{\mathcal C}\, n_{\mathbf{i}}(t)\, n_{\mathbf{j}}(t') \bigr\rangle - 1\Bigr],\\
\chi^s_{\mathbf{ij}}(t,t') &\simeq -\tfrac{\ii}{2}\bigl\langle \mathcal T_{\mathcal C}\, S^z_{\mathbf{i}}(t)\, S^z_{\mathbf{j}}(t') \bigr\rangle,
\end{aligned}
\label{eq:correlators}
\end{equation}
with $n_{\mathbf{i}} \equiv n_{\mathbf{i}\uparrow}+n_{\mathbf{i}\downarrow}$ and $S^z_{\mathbf{i}} \equiv n_{\mathbf{i}\uparrow}-n_{\mathbf{i}\downarrow}$. The factor $\tfrac12$ makes $\chi^c$ and $\chi^s$ responses per spin species, so that both reduce to the bubble $\chi$ at $U=0$.
FLEX is a conserving approximation~\cite{Bickers1989}, so that particle number is conserved and the total energy changes only through the work done by the interaction ramp.
Note that with the above definition, our susceptibilities have the opposite sign to those of the original FLEX formulation by Bickers and Scalapino~\cite{Bickers1989b}.
Diagrammatically, the FLEX self-energy, Eq.~\eqref{eq:sigma_flex}, can be represented as~\cite{Schmalian1996,Geng2025}
\begin{equation}
\vcenter{\hbox{\includegraphics[width=0.85\columnwidth]{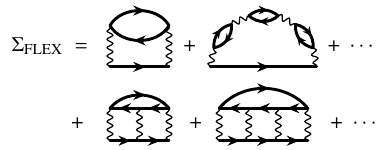}}}
\label{eq:flex_diagrams}
\end{equation}
where thick solid lines with arrows denote the interacting $G(t,t')$ and wiggly lines $U(t)\,\delta_{\mathcal C}(t,t')$.
The Hartree diagram vanishes at half filling~\cite{Aoki2014}.
As in Refs.~\cite{Schmalian1996,Shinaoka2023,Kitatani2015}, we neglect the particle--particle channel, which is much smaller than the spin-fluctuation contribution near half filling in equilibrium~\cite{Bickers1989}.

At half filling, FLEX describes a paramagnetic equilibrium state at high temperature and an AFM instability at low temperature. The instability occurs when $\chi^s_{\mathbf{q}}$ diverges at $\mathbf{q}=(\pi,\pi)$, the nesting vector of the half-filled Fermi surface.
The Mermin--Wagner theorem forbids AFM long-range order at finite temperature in two dimensions for short-range interactions and a continuous symmetry (here the spin rotation symmetry)~\cite{Mermin1966,Ghosh1971}. Self-consistent FLEX is expected to obey the theorem~\cite{Kontani2006,Kitatani2015}, and no instability is found on large momentum grids~\cite{Deisz1996}.
The regime
with strongly enhanced correlations 
nevertheless remains of interest, since in three dimensions, or already with a weak interlayer coupling, the AFM phase exists at nonzero temperature~\cite{Bickers1989,Takimoto2002}. FLEX studies of two-dimensional systems therefore often define an AFM phase boundary by an arbitrary threshold on the static denominator, $1+U\chi_{(\pi,\pi)}(\omega=0)$, or on $\chi^s$ itself~\cite{Kitatani2015,Sakurazawa2005,Hatakeyama2014,Nakamura2026,Oka2010}. In our calculations, spin symmetry is enforced, so that long-range AFM order cannot develop and results inside the AFM phase would be unphysical, but $\chi^s$ can still grow strongly as the instability is approached. The dynamics is meaningful provided that the correlation length remains well below the system size. 

\subsection{Migdal approximation}
For the electron--phonon model, we use the self-consistent Migdal approximation~\cite{Migdal1958,Murakami2015,Abdurazakov2018}, as in Ref.~\cite{Sroda2026}. In addition to $G$, this diagrammatic scheme involves the phonon propagator
\begin{equation}
D_{\mathbf{q}}(t,t') = -\ii \bigl\langle \mathcal{T}_{\mathcal C}\, X_{\mathbf{q}}(t)\, X^\dagger_{\mathbf{q}}(t') \bigr\rangle,
\end{equation}
defined in terms of the displacement $X_{\mathbf{q}} = b^\dagger_{-\mathbf{q}} + b_{\mathbf{q}}$. The electron self-energy is~\cite{Stefanucci2025}
\begin{equation}
\Sigma_{\mathbf{ij}}(t,t') = \vcenter{\hbox{\includegraphics[scale=1.1,page=1]{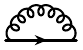}}} = \ii\,G_{\mathbf{ij}}(t,t')\,V_{\mathbf{ij}}(t,t'),
\label{eq:migdal_diagrams}
\end{equation}
where the coiled line denotes the phonon-mediated interaction $V_{\mathbf{q}}(t,t') = g_{\mathbf{q}}(t)\,D_{\mathbf{q}}(t,t')\,g_{\mathbf{q}}(t')$. The phonon propagator is obtained from the bosonic Dyson equation
\begin{equation}
D_{\mathbf{q}} = D^0_{\mathbf{q}} + D^0_{\mathbf{q}} * \Pi_{\mathbf{q}} * D_{\mathbf{q}},
\label{eq:phonon_dyson}
\end{equation}
where $D^0_{\mathbf{q}}$ is the bare propagator and $\Pi_{\mathbf{q}}(t,t') = 2\,g_{\mathbf{q}}(t)\,\chi_{\mathbf{q}}(t,t')\,g_{\mathbf{q}}(t')$ is the phonon self-energy, with the factor 2 coming from the spin sum in the electron--phonon coupling term, Eq.~\eqref{eq:H_eph}, and with the polarization bubble $\chi_{\mathbf{q}}$ defined above. The electron Green's function and the phonon propagator are determined self-consistently from the same functional, which makes the approximation conserving~\cite{Baym1962,Stefanucci2025}.
For the analysis, we compute the charge susceptibility from the converged solution,
\begin{equation}
\chi^c_{\mathbf{q}} = \chi_{\mathbf{q}} + \chi_{\mathbf{q}} * 2V^0_{\mathbf{q}} * \chi^c_{\mathbf{q}},
\end{equation}
with the bare interaction $V^0_{\mathbf{q}}(t,t') = g_{\mathbf{q}}(t)\,D^0_{\mathbf{q}}(t,t')\,g_{\mathbf{q}}(t')$. It approximates the same charge correlation function as in FLEX, Eq.~\eqref{eq:correlators}.
At half filling, the model has a CDW instability, which occurs when $\chi^c_{\mathbf{q}}$ diverges at the nesting vector $\mathbf{q}=(\pi,\pi)$~\cite{Sroda2026}. Since our calculations assume a one-site unit cell, CDW order cannot develop.

\subsection{Numerical implementation}
On a grid of $N_t$ time steps, a two-time function costs $O(N_t^2)$ memory per $\mathbf{k}$ point, and solving the Dyson equation costs $O(N_t^3)$ operations~\cite{Aoki2014,Schuler2019}. This limits conventional calculations to short times and coarse momentum grids.
We overcome this limitation by using the recently developed NEGF framework based on quantics tensor trains (QTTs)~\cite{Shinaoka2023,Sroda2025,Inayoshi2025,Sroda2025b}. The time arguments are written in a binary form, $t \rightarrow (T_1, T_2, \dots, T_{N_b})$ with bits $T_n \in \{0,1\}$ and $N_b = \log_2 N_t$. The bits of $t$ and $t'$ are interleaved, and, for each momentum $\mathbf{k}$, the resulting tensor is expressed as a tensor train, or matrix product state:
\begin{equation}
\vcenter{\hbox{\includegraphics[width=0.7\columnwidth]{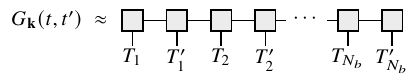}}}.
\label{eq:qtt}
\end{equation}
Here, each square represents a tensor carrying one bit, and neighboring tensors are connected by bonds of dimension at most $M$, the bond dimension.
For bounded $M$, the memory cost $O(N_b M^2)$ grows only linearly with $N_b$, i.e., logarithmically with $N_t$~\cite{Shinaoka2023}.

In both models, the self-energies depend on $G$, so that the Dyson equations are solved self-consistently. In each iteration, the self-energies and polarizations are evaluated in real space, while the Dyson equations are solved in momentum space. The iterations start from $G=G^0$, together with $\chi^c=\chi^s=\chi$ in FLEX and $D=D^0$ in the Migdal approximation. Using the Langreth rules~\cite{Aoki2014,Schuler2019}, the Dyson equation is split into four coupled equations for its physical components (Matsubara, left-mixing, lesser $G^<$, and retarded $G^R$)~\cite{Sroda2025}. Long times are reached in block time steps~\cite{Inayoshi2025}: the solution is extrapolated to the next time block, and this prediction serves as the starting point of the self-consistency loop on that block~\cite{Sroda2025b}.

Since iterations on large momentum grids are expensive, we first propagate the self-consistent solution with block time steps up to $t_{\max}$ for a momentum grid of $N_k = 32\times32$ points. We then interpolate the self-energy in momentum onto a grid with twice as many points per direction and iterate the self-consistency loop again, now over the whole contour at once, and we repeat this up to $N_k = 256\times256$, similar to Ref.~\cite{Sroda2026}. In this way, most iterations are performed on the small grids.
At $N_k = 256\times256$, a single two-time function remains large even in compressed form: it takes of the order of 15\,GB in momentum space (irreducible zone) and 30--60\,GB in real space. An iteration of the self-consistency loop requires several hundred GB, close to the memory of a single computational node. We therefore stop at this grid size. Larger grids would require compressing the momentum dependence as well.
The implementation is written in Julia~\cite{bezanson2017julia} and is based on the ITensor library~\cite{itensor} and libraries developed by the tensor4all group~\cite{Fernandez2024,tensor4all}.

\subsection{Effective temperature and correlation length}
We characterize the state after the ramp by measuring effective inverse temperatures, defined in two ways. The first is energy matching: the inverse temperature is taken from an equilibrium simulation with the same parameters and the same total energy as the state after the ramp.
The second is a fit of distribution functions. We Fourier transform local two-time functions $X$ with respect to the relative time $t_{\mathrm{rel}}=t-t'$ at fixed average time $t_{\mathrm{av}}=(t+t')/2$, and fit
\begin{equation}
f_X(\omega,t_{\mathrm{av}}) = \mp\frac{\mathrm{Im}\,X^<(\omega,t_{\mathrm{av}})}{2\,\mathrm{Im}\,X^R(\omega,t_{\mathrm{av}})}
\label{eq:distribution}
\end{equation}
to $1/(e^{\beta\omega}\pm1)$, where $X^<$ and $X^R$ are the lesser and retarded components of $X$. For the fermionic functions, $G$ and $\Sigma$, the upper signs apply and the fit function is the Fermi--Dirac distribution. For the bosonic functions, $\chi^c$ and $\chi^s$ in the Hubbard model and $D$ and $\chi^c$ in the electron--phonon model, the lower signs apply and the fit function is the Bose--Einstein distribution.

In addition, we extract a spin [charge] correlation length $\xi$ from the equal-time correlations in momentum space, $\ii\,\chi^{s,<}_{\mathbf q}(t,t)$ [$\ii\,\chi^{c,<}_{\mathbf q}(t,t)$], for the Hubbard [electron--phonon] model. Close to the instability, this function has a peak at $\mathbf q = (\pi,\pi)$, which becomes narrower as the correlations extend further. In equilibrium calculations, $\xi$ is obtained from a fit of that peak to the Ornstein--Zernike form $A/(\xi^{-2}+\delta q^2)$~\cite{LeBlanc2019,Schafer2021}. This Lorentzian falls to half of its maximum at a distance $\delta q = 1/\xi$ from $(\pi,\pi)$. In the nonequilibrium case, we therefore define
\begin{equation}
\xi(t) = \frac{1}{\kappa(t)},
\label{eq:xi_fit}
\end{equation}
where $\kappa(t)$ is the half width at half maximum of the peak along a lattice axis. Before $\kappa$ is determined, we remove from these correlations their on-site contribution (the $\mathbf r = 0$ term of the Fourier sum, equal to the momentum average), a momentum-independent background which would otherwise shift the peak upward and increase the measured half width.
An advantage of this definition is that it does not involve any fit. In contrast, when one tries to fit the decay of the equal-time correlations in real space, the resulting $\xi$ depends strongly on the choice of the fit window.

\section{Results and Discussion}\label{results}

\begin{figure}[tb]
	\centering
	\includegraphics[width=\columnwidth]{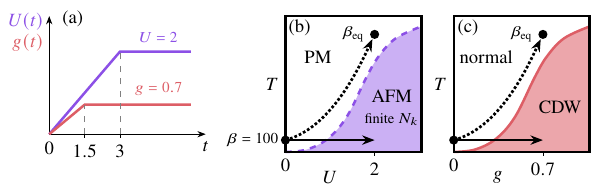}
	\caption{Ramp protocol and schematic phase diagrams. (a) Ramps of $U$ in the Hubbard model and of $g$ in the electron--phonon model. (b) Phase diagram of the Hubbard model within FLEX, with the paramagnetic (PM) phase and the region of strong AFM correlations beyond the finite-grid instability (dashed line). (c) Phase diagram of the electron--phonon model, with the normal and CDW phases separated by a finite-temperature transition~\cite{Weber2018}. In (b) and (c), the solid arrow shows the ramp in the absence of heating, and the dotted arrow qualitatively sketches the evolution of the system to the thermalized state at $\beta_{\mathrm{eq}}$. Note that during the evolution, the state is not thermal, so that a global temperature cannot be defined.}
	\label{fig_protocol}
\end{figure}

\subsection{Setup}\label{sec:setup}

The energy unit is the hopping $t_h$ and the time unit is $1/t_h$, with $\hbar = k_B = 1$. For both models, the largest momentum grid has $N_k = 256 \times 256$ points, and the real-time axis is discretized with $2^{29}$ points up to $t_{\max} = 80$ for the Hubbard model and $2^{30}$ points up to $t_{\max} = 100$ for the electron--phonon model. The tensor trains are truncated with a relative cutoff $\tau = 10^{-10}$ ($10^{-9}$), defined by $\lVert f-\tilde f\rVert_F^2/\lVert f\rVert_F^2 < \tau$ for a function $f$ and its compressed form $\tilde f$~\cite{Sroda2026}, and a maximal bond dimension $M = 60$ ($65$) for the Hubbard (electron--phonon) model. The convergence of the self-consistency loop is monitored by $\epsilon = \max_{\mathbf k}\sum_{c}\lVert G^{c}_{\mathbf k,\mathrm{new}}-G^{c}_{\mathbf k,\mathrm{old}}\rVert_F/\lVert G^{c}_{\mathbf k,\mathrm{old}}\rVert_F$, where $G_{\mathrm{old}}$ and $G_{\mathrm{new}}$ are the Green's functions before and after an iteration, and the sum runs over the retarded, left-mixing, and lesser components~\cite{Sroda2026}. For both models, we reach $\epsilon < 10^{-3}$ on the $32\times32$ and $64\times64$ grids, and $\epsilon \approx 10^{-2}$ on the $128\times128$ and $256\times256$ grids.

The calculations start from the noninteracting thermal state at inverse temperature $\beta = 100$, and the interaction is then switched on by a linear ramp [Fig.~\ref{fig_protocol}(a)]. In the Hubbard model, $U$ increases to 2 over three time units, and in the electron--phonon model, with a maximal acoustic-phonon energy $\Omega_0 = 1$, $g$ increases to 0.7 over 1.5 time units. We use ramps rather than quenches to limit the heating of the system, as sketched in Fig.~\ref{fig_protocol}(b,c). After the ramp, the total energy is conserved. For the Hubbard model, energy matching and the fits of the distribution functions [Eq.~\eqref{eq:distribution}] at long times both give $\beta_{\mathrm{eq}} \approx 10.3$. For the electron--phonon model, energy matching and the fits of the electronic distribution functions give $\beta_{\mathrm{eq}} \approx 8.0$, while the phonons have not yet reached this temperature at the longest simulated time (Sec.~\ref{sec:temperature}).

\begin{figure*}[t]
	\begin{minipage}[t]{0.49\textwidth}
		\small (a) Hubbard model\\[2pt]
		\includegraphics[width=\linewidth]{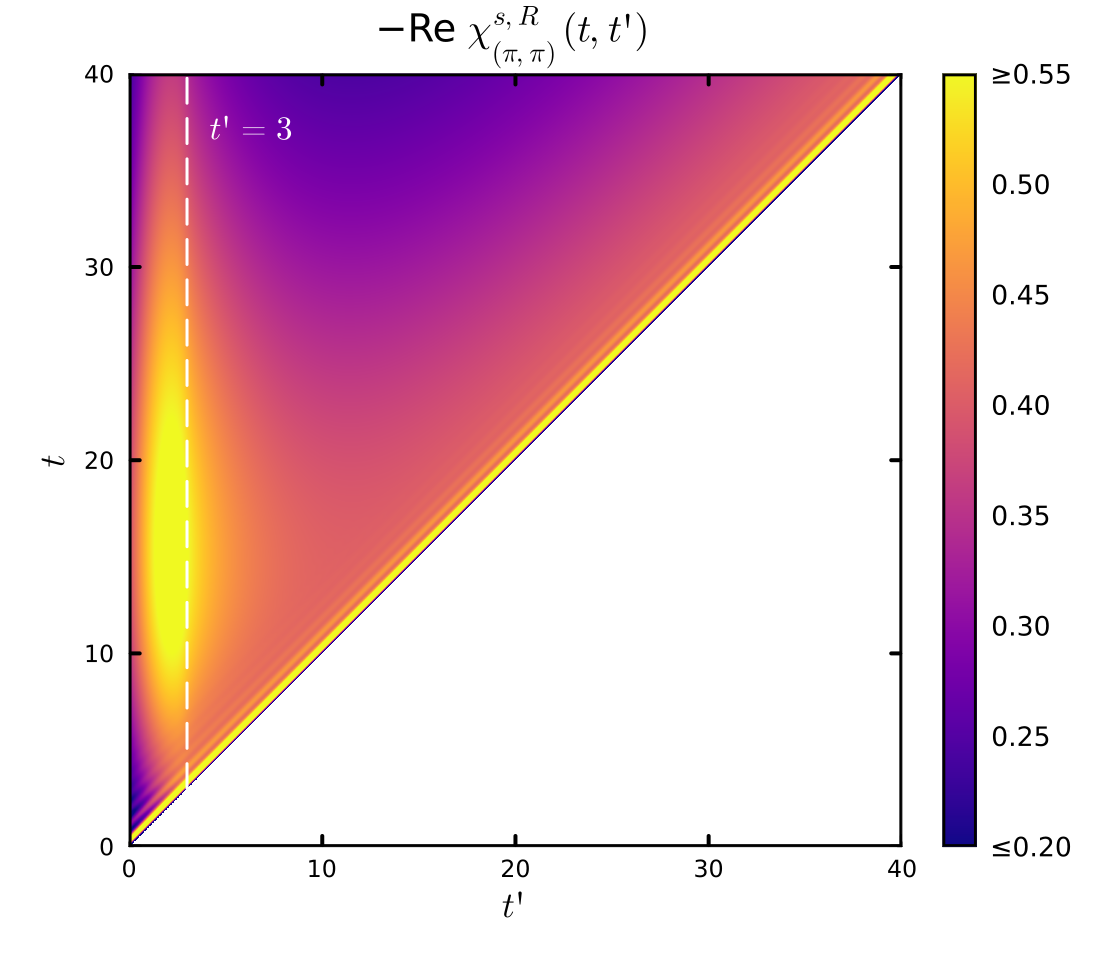}
	\end{minipage}\hfill
	\begin{minipage}[t]{0.49\textwidth}
		\small (b) Electron--phonon model\\[2pt]
		\includegraphics[width=\linewidth]{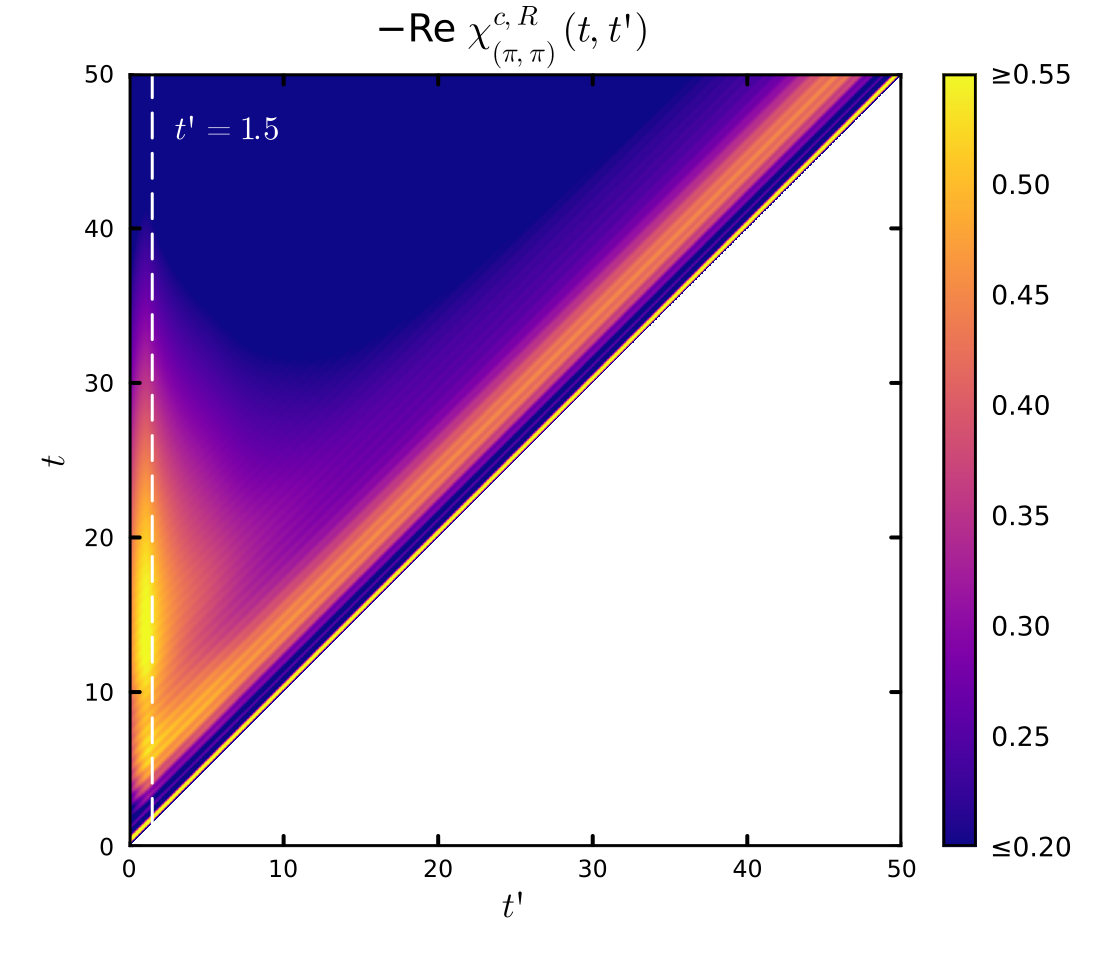}
	\end{minipage}
	\caption{(a) Retarded spin susceptibility $-\mathrm{Re}\,\chi^{s,R}_{(\pi,\pi)}(t,t')$ of the Hubbard model and (b) retarded charge susceptibility $-\mathrm{Re}\,\chi^{c,R}_{(\pi,\pi)}(t,t')$ of the electron--phonon model, both for $N_k = 256 \times 256$. The vertical dashed lines mark the end of the ramp.}
	\label{fig_transient_order}
\end{figure*}

\begin{figure*}[t]
	\begin{minipage}[t]{0.49\textwidth}
		\small (a) Hubbard model\\[2pt]
		\includegraphics[width=\linewidth]{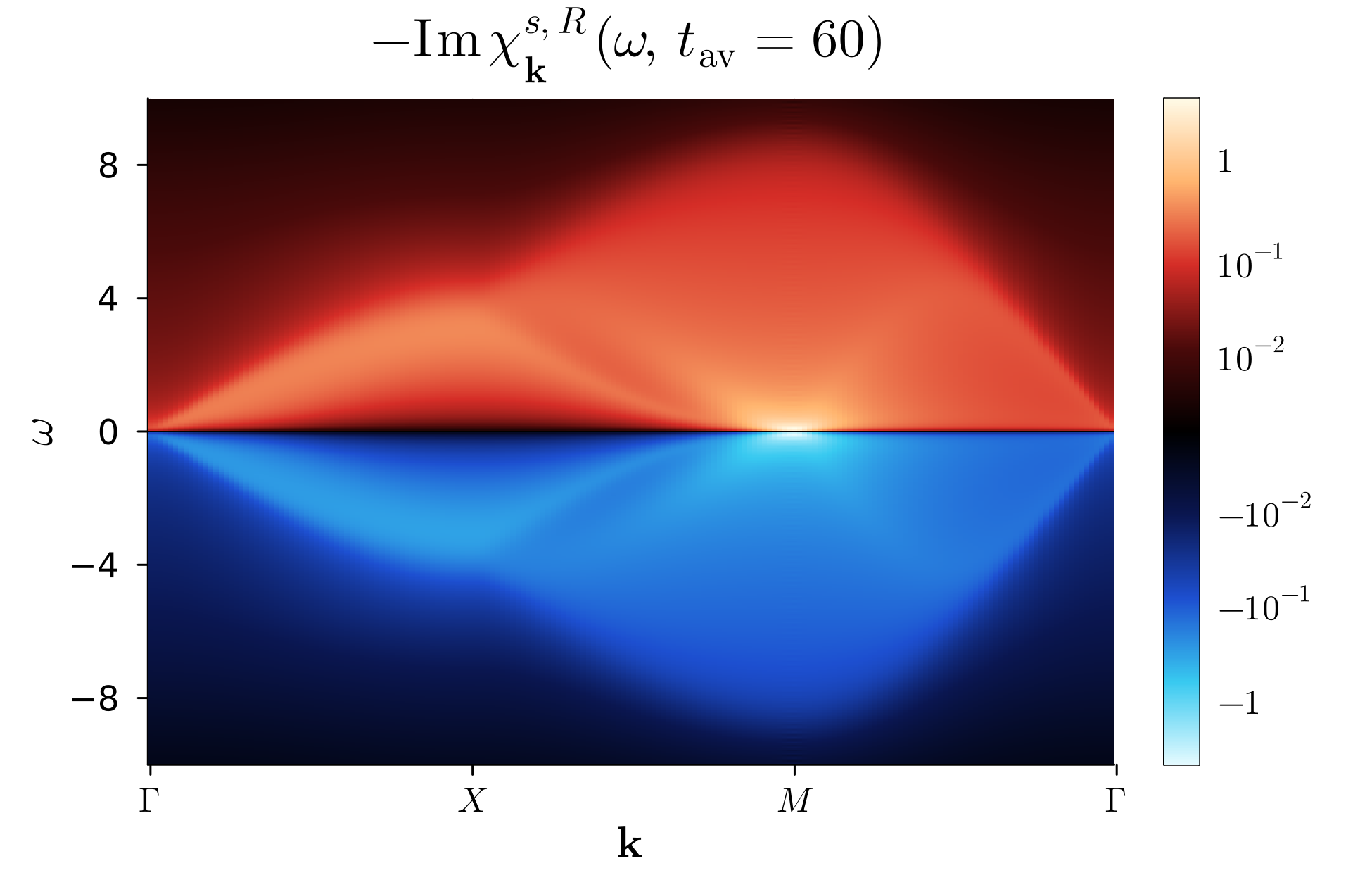}
	\end{minipage}\hfill
	\begin{minipage}[t]{0.49\textwidth}
		\small (b) Electron--phonon model\\[2pt]
		\includegraphics[width=\linewidth]{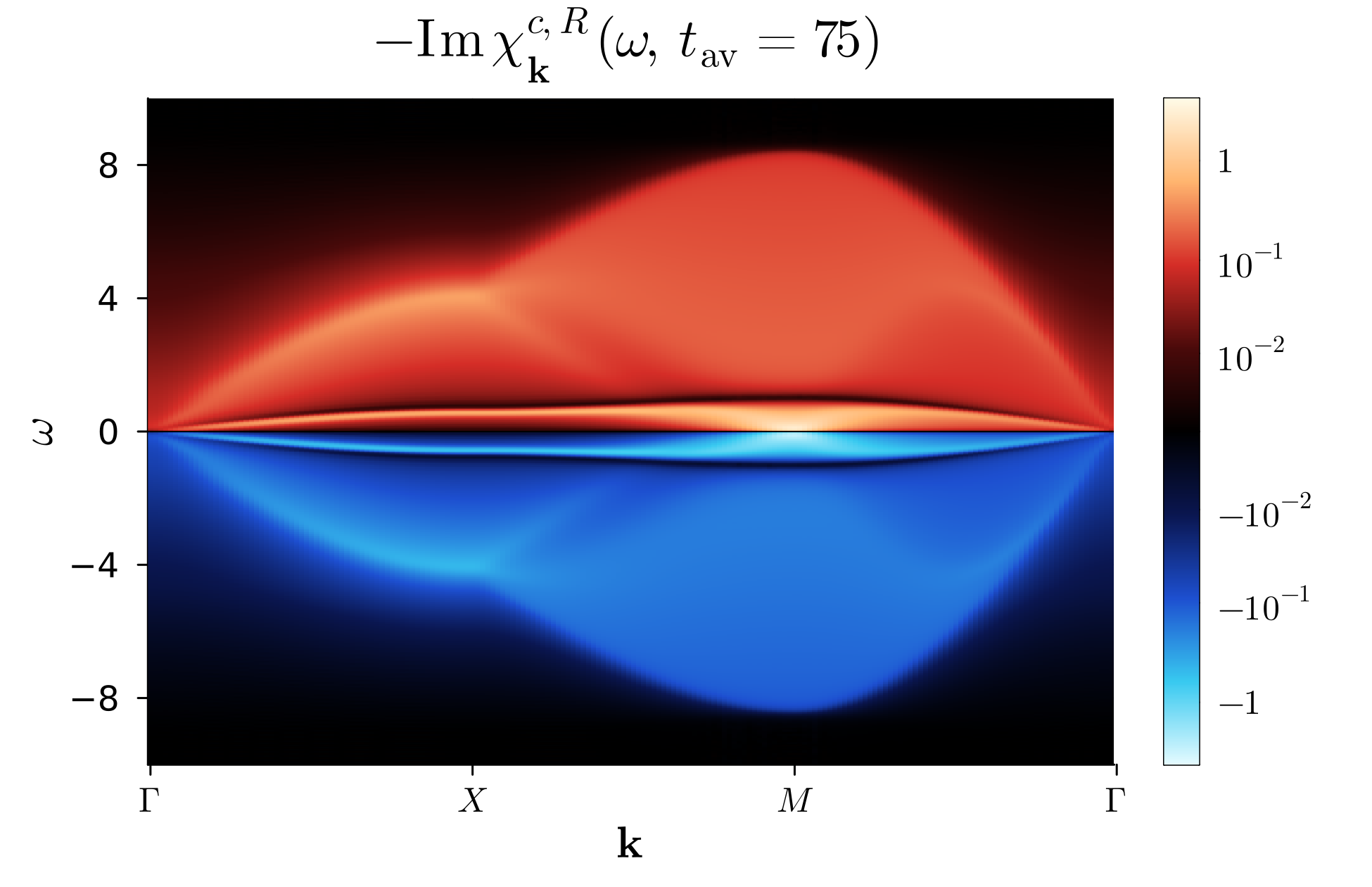}
	\end{minipage}
	\caption{Spectra in the late, nearly thermal, state after the ramp, along $\Gamma$--$X$--$M$--$\Gamma$ for $N_k = 128\times128$, on a symmetric logarithmic color scale. They are obtained from the Fourier transform with respect to $t_{\mathrm{rel}}$ with a Gaussian window of width 20, which suppresses ringing from the slow decay near $M$. (a) Spin spectrum $-\mathrm{Im}\,\chi^{s,R}_{\mathbf k}(\omega, t_{\mathrm{av}})$ of the Hubbard model at $t_{\mathrm{av}} = 60$. (b) Charge spectrum $-\mathrm{Im}\,\chi^{c,R}_{\mathbf k}(\omega, t_{\mathrm{av}})$ of the electron--phonon model at $t_{\mathrm{av}} = 75$. The peak at $M$ reaches 4.18 at $\omega \approx 0.08$ (a) and 2.77 at $\omega \approx 0.09$ (b). The low-energy branch in (b) is the renormalized phonon, see Ref.~\cite{Sroda2026}.}
	\label{fig_spectra}
\end{figure*}

\subsection{Transient enhancement of the antiferromagnetic and charge-density-wave responses}\label{sec:transient}
Figure~\ref{fig_transient_order} shows the retarded susceptibilities at the nesting vector $(\pi,\pi)$: (a) the spin susceptibility $-\mathrm{Re}\,\chi^{s,R}_{(\pi,\pi)}(t,t')$ of the Hubbard model and (b) the charge susceptibility $-\mathrm{Re}\,\chi^{c,R}_{(\pi,\pi)}(t,t')$ of the electron--phonon model. The retarded component is the linear-response function~\cite{Stefanucci2025}. The staggered magnetization $\delta m$ induced by a weak staggered field $\delta h$ is $\delta m(t) = -2\int_0^t dt'\,\chi^{s,R}_{(\pi,\pi)}(t,t')\,\delta h(t')$, and similarly for the charge response to a staggered potential.
The minus sign compensates for our sign convention [Eq.~\eqref{eq:correlators}], so that positive values correspond to an AFM or CDW response. Since the vertex corrections missing in $\chi^s$ are expected to reduce it~\cite{Monthoux1994}, the magnitude of the enhancement discussed below is likely overestimated.

\paragraph*{Hubbard model.}
Some features of the response in Fig.~\ref{fig_transient_order}(a) depend only on $t-t'$ and are present for all perturbation times. Close to the diagonal, the sharp peak at $t-t' \approx 0.3$ (seen as a yellow line) is the immediate response of the electrons to the perturbation. It is followed by oscillations in $t-t'$ with period $2\pi/(8t_h) \approx 0.8$, i.e., a frequency close to the largest energy of a particle--hole pair with momentum $(\pi,\pi)$, the bandwidth $8t_h$ [Fig.~\ref{fig_spectra}(a)]. Both features come from particle--hole excitations, and are already present in the noninteracting bubble at $(\pi,\pi)$.

\begin{figure*}[t]
	\begin{minipage}[t]{0.49\textwidth}
		\small (a) Hubbard model\\[2pt]
		\includegraphics[width=\linewidth]{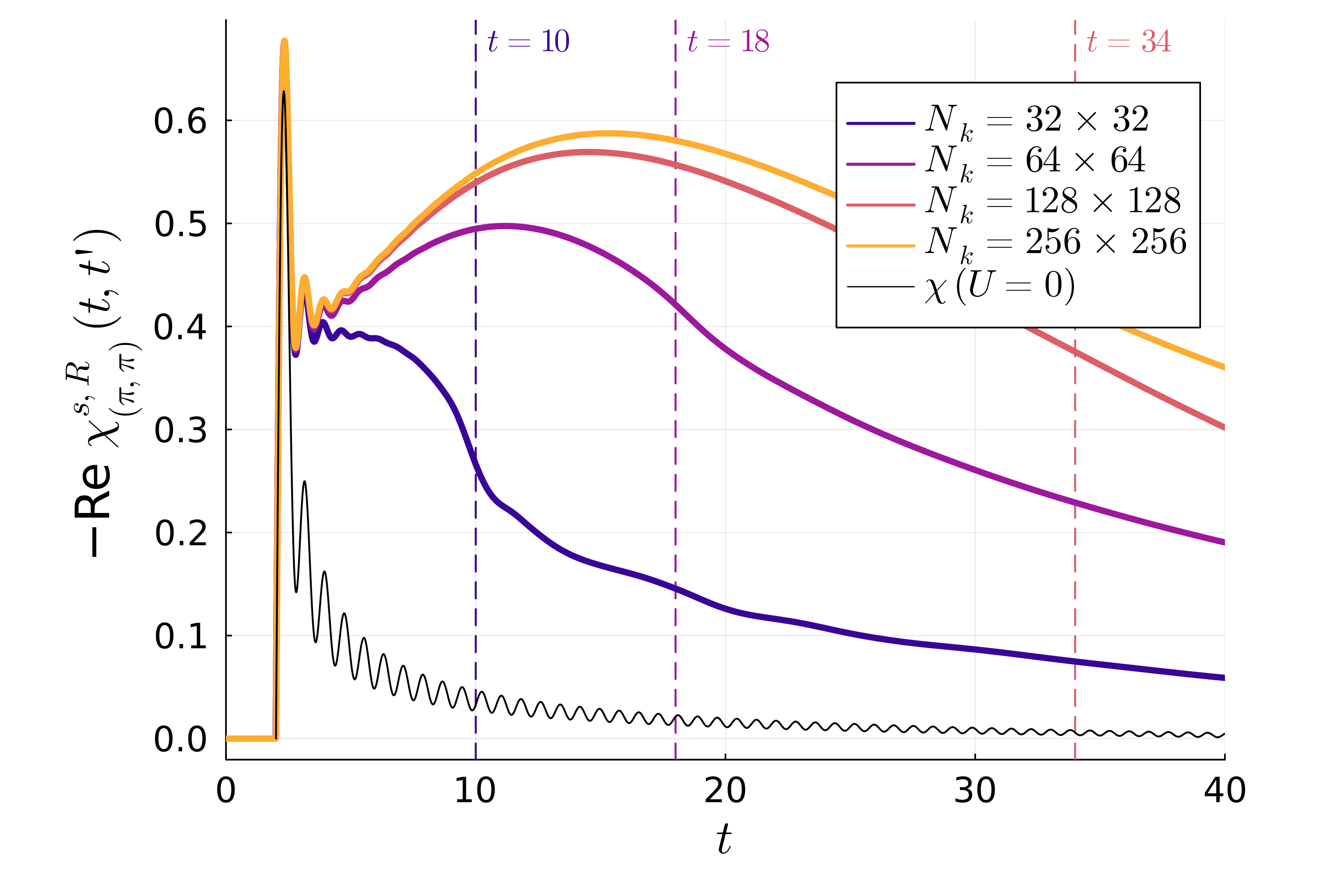}
	\end{minipage}\hfill
	\begin{minipage}[t]{0.49\textwidth}
		\small (b) Electron--phonon model\\[2pt]
		\includegraphics[width=\linewidth]{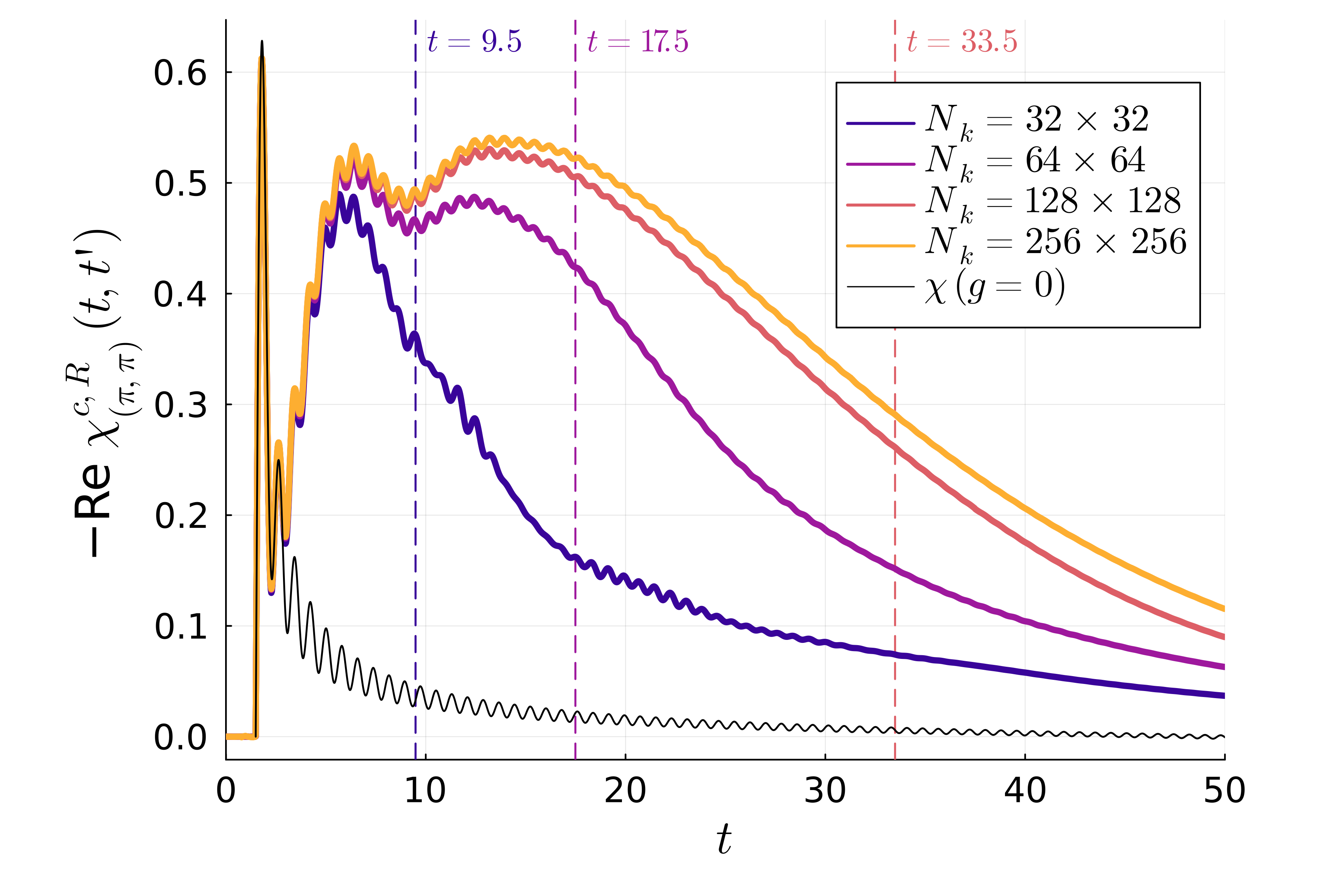}
	\end{minipage}
	\caption{Cuts of (a) $-\mathrm{Re}\,\chi^{s,R}_{(\pi,\pi)}(t,t')$ of the Hubbard model at $t'=2$ and (b) $-\mathrm{Re}\,\chi^{c,R}_{(\pi,\pi)}(t,t')$ of the electron--phonon model at $t' = 1.5$, as a function of $t$ for different lattice sizes. The thin black lines show the same cuts for the noninteracting bubble. The dashed vertical lines mark the times $t' + L/4$ at which the electronic fronts meet and interfere on the periodic $L\times L$ lattice, for $L = 32$, 64, and 128 (see text).
	}
	\label{fig_finite_size}
\end{figure*}

In contrast, the transient part of the response, which depends on the perturbation time $t'$, is a slow component. For perturbations during and shortly after the ramp, the response keeps growing with $t$, reaches its maximum around $t \approx 15$, and then decays [see also the cut at $t' = 2$ in Fig.~\ref{fig_finite_size}(a)]. For later perturbation times, the slow component becomes weaker and gradually stops growing. For $t' \gtrsim 20$, the response depends only on $t-t'$, i.e., the system has reached a stationary, equilibrium-like regime. Measured at the same time delay $t-t'$, the maximal response to a perturbation during the ramp is about 1.8 times larger than in the stationary regime.

The slow component revealed by the simulations is the signature of a soft collective mode at $(\pi,\pi)$, a paramagnon, i.e., an antiferromagnetic spin fluctuation of the paramagnetic phase, which appears close to the instability to the AFM phase. In the nearly thermal state after the ramp, the spin spectrum shows a sharp peak at $(\pi,\pi)$ and $\omega \approx 0$, much larger than the weak particle--hole continuum [Fig.~\ref{fig_spectra}(a); note the logarithmic scale]. The mode is soft: it spans a narrow energy range around $\omega = 0$
and hence 
 decays over tens of time units, much more slowly than the particle--hole excitations.
Its enhancement after the ramp can be understood from the Dyson equation for $\chi^s$, $\chi^s = \chi/(1+U\chi)$, which would diverge at the instability $1+U\chi_{(\pi,\pi)} = 0$~\cite{Bickers1989} (in our convention, $\chi_{(\pi,\pi)} < 0$).
The denominator depends on $U$ and on the bubble, and the two change on different time scales. $U$ reaches its final value at the end of the ramp, when the electrons have not yet fully heated up, so that the denominator is transiently smaller than its thermalized value. Indeed, $|\chi_{(\pi,\pi)}|$ is larger at lower temperature and decreases as the electrons heat up. Since the bubble is built from $G$, it follows the electron distribution, which heats up
with some delay. 
As shown in Sec.~\ref{sec:temperature}, the electrons remain colder than the thermalized state long after the end of the ramp.
To quantify how small the denominator becomes, we evaluate its static value $1+U\chi_{(\pi,\pi)}(t,\omega=0)$, with $\chi_{(\pi,\pi)}(t,\omega=0) = \int_0^t dt'\,\chi^R_{(\pi,\pi)}(t,t')$. Between $t \approx 3$ and $t \approx 18$, the denominator even changes sign, from positive to slightly negative, with a minimum of about $-0.09$ at $t \approx 7$, so that the paramagnetic state is transiently unstable and the response to a perturbation applied during or shortly after the ramp grows instead of decaying [Figs.~\ref{fig_transient_order}(a) and \ref{fig_finite_size}(a)]. As the electrons heat up, the denominator becomes positive again and approaches about $0.03$, and the response to later perturbations decays.
In summary, the transient enhancement requires two ingredients: a final state close to the instability, and a ramp faster than the heating of the electrons, so that $U$ reaches its final value while the electrons are still cold and $|\chi_{(\pi,\pi)}|$ is still large.

The transient instability, followed by a stable paramagnetic final state, corresponds to the transient short-range order regime discussed by Bauer \textit{et al.}~\cite{Bauer2015}, which was reported for different ramp parameters. They identified this regime from the equal-time spin correlation, which in our calculations rises after the ramp and then decays, as does the correlation length (Sec.~\ref{sec:xi}). However, for the same ramp, to $U = 2$ within three time units, they found exponentially growing correlations and an ordered final state. A likely reason for this difference is the self-energy. Bauer \textit{et al.} used a local second-order self-energy, which does not incorporate spin fluctuations. In FLEX, in contrast, $\Sigma$ contains $\chi^s$, which closes a feedback loop: as $\chi^s$ grows near $(\pi,\pi)$, $\Sigma$ increases the scattering rate of the electrons near the Fermi surface, so that $G$ decays faster in $t-t'$ for these electrons, and so does the bubble at $(\pi,\pi)$, to which they contribute most. As a result, the bubble $|\chi_{(\pi,\pi)}|$ decreases and the denominator $1+U\chi_{(\pi,\pi)}$ moves away from zero, which limits the growth of $\chi^s$~\cite{Gukelberger2015}. In equilibrium, this feedback prevents the AFM transition in the thermodynamic limit~\cite{Deisz1996,Kontani2006}. In our FLEX calculations, the instability therefore remains weak and ends when the electrons heat up, so that the response grows only moderately and the final state is paramagnetic. A treatment without this feedback likely overestimates the enhancement.
Dasari and Eckstein~\cite{Dasari2019}, using nonequilibrium FLEX, reported a slow $(\pi,\pi)$ mode in the initial equilibrium state, which is quickly suppressed by a strong electric field pulse. Here, in contrast, it is transiently excited by the ramp.

\begin{figure*}[t]
	\begin{minipage}[t]{0.49\textwidth}
		\small (a) Hubbard model\\[2pt]
		\includegraphics[width=\linewidth]{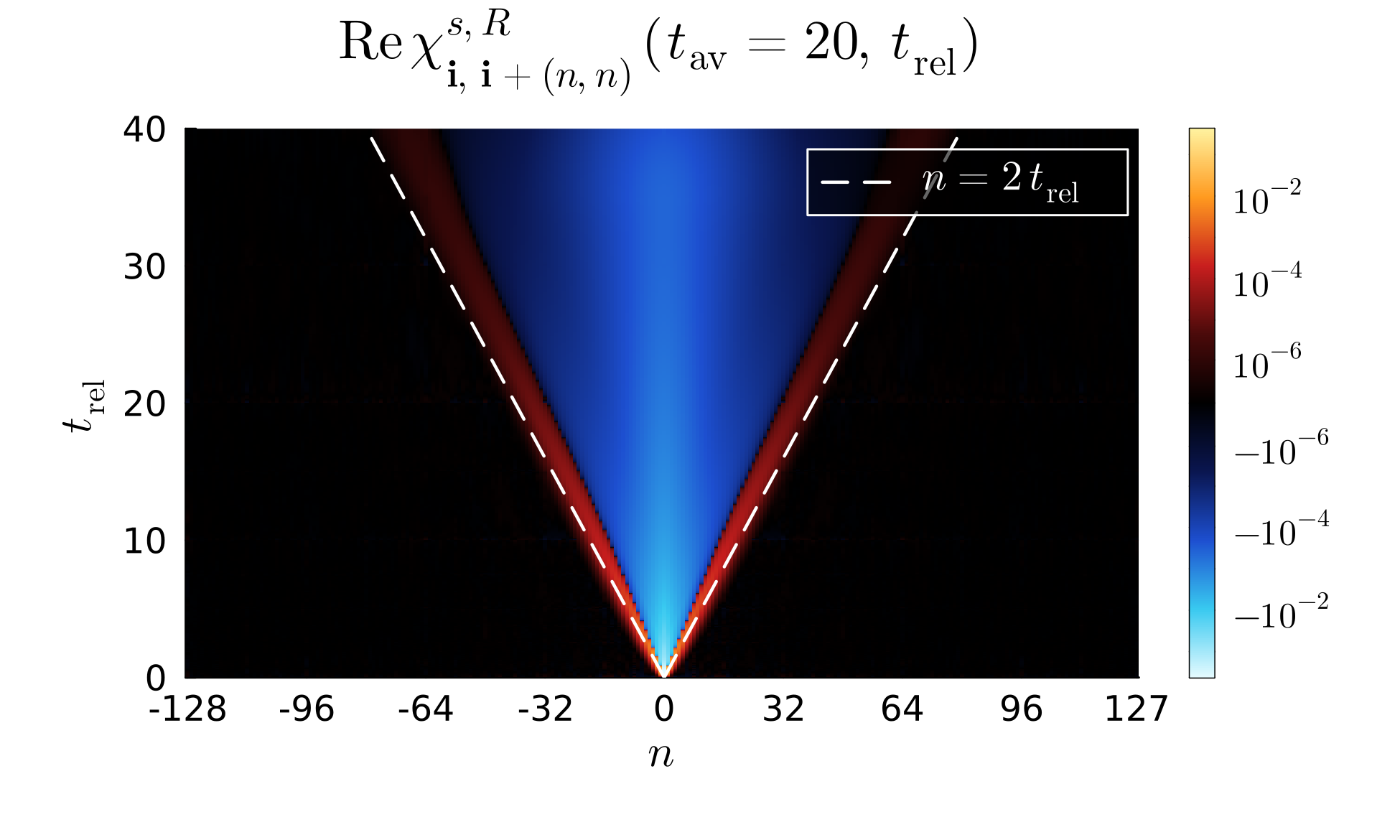}
	\end{minipage}\hfill
	\begin{minipage}[t]{0.49\textwidth}
		\small (b) Electron--phonon model\\[2pt]
		\includegraphics[width=\linewidth]{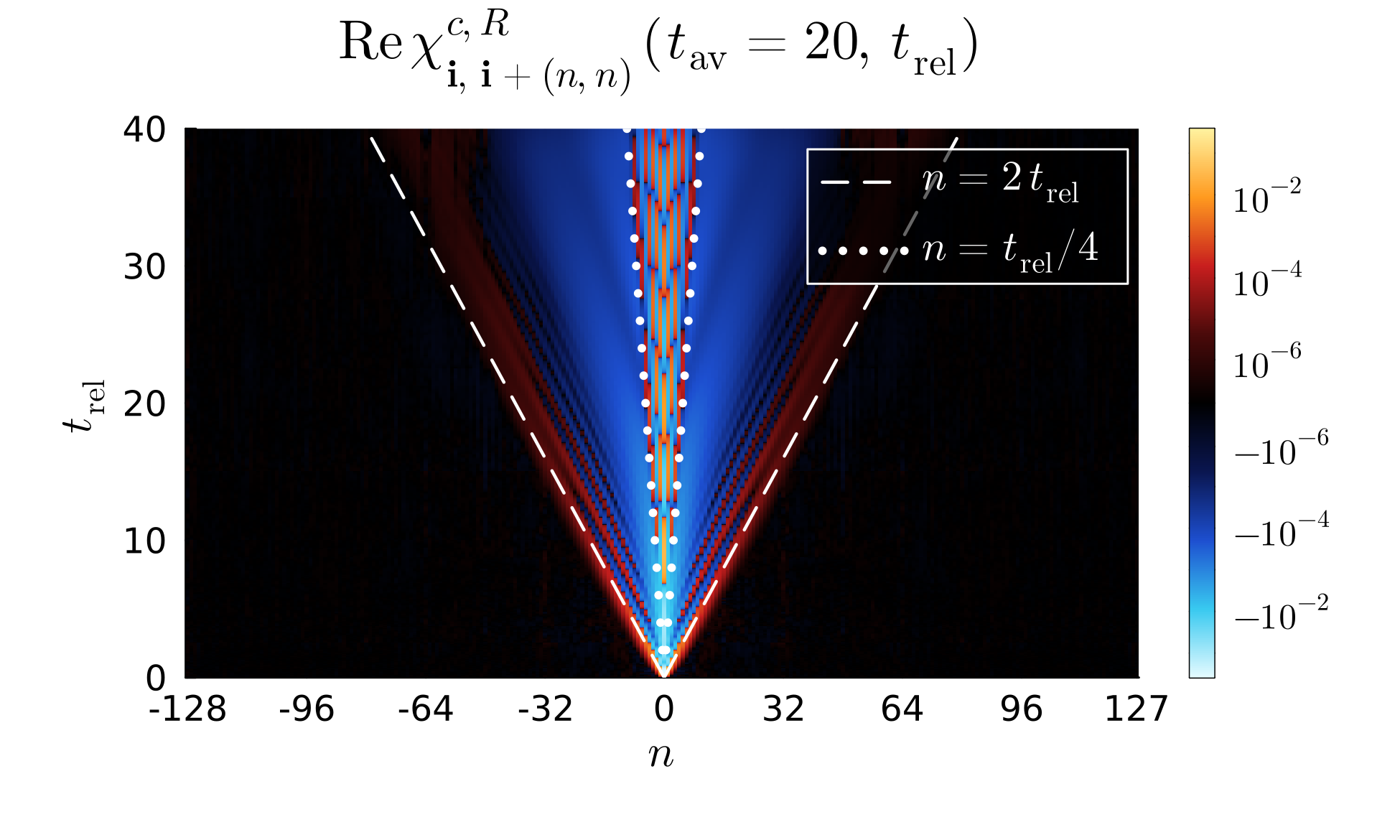}
	\end{minipage}
	\caption{Real-space light cone along the diagonal, $\mathbf{j} = \mathbf{i} + (n,n)$, as a function of $t_\mathrm{rel}$, on a symmetric logarithmic color scale. (a) $\mathrm{Re}\,\chi^{s,R}_{\mathbf{i},\mathbf{i}+(n,n)}(t_\mathrm{av}, t_\mathrm{rel})$ of the Hubbard model at $t_\mathrm{av} = 20$ for $N_k = 256 \times 256$. (b) $\mathrm{Re}\,\chi^{c,R}_{\mathbf{i},\mathbf{i}+(n,n)}(t_\mathrm{av}, t_\mathrm{rel})$ of the electron--phonon model at $t_\mathrm{av} = 20$ for $N_k = 256 \times 256$. The white dashed lines correspond to the maximal group velocity of the noninteracting electrons, and the white dotted lines in (b) to that of the bare phonons.}
	\label{fig_cone}
\end{figure*}

\begin{figure*}[t]
	\begin{minipage}[t]{0.49\textwidth}
		\small (a) Hubbard model\\[2pt]
		\includegraphics[width=\linewidth]{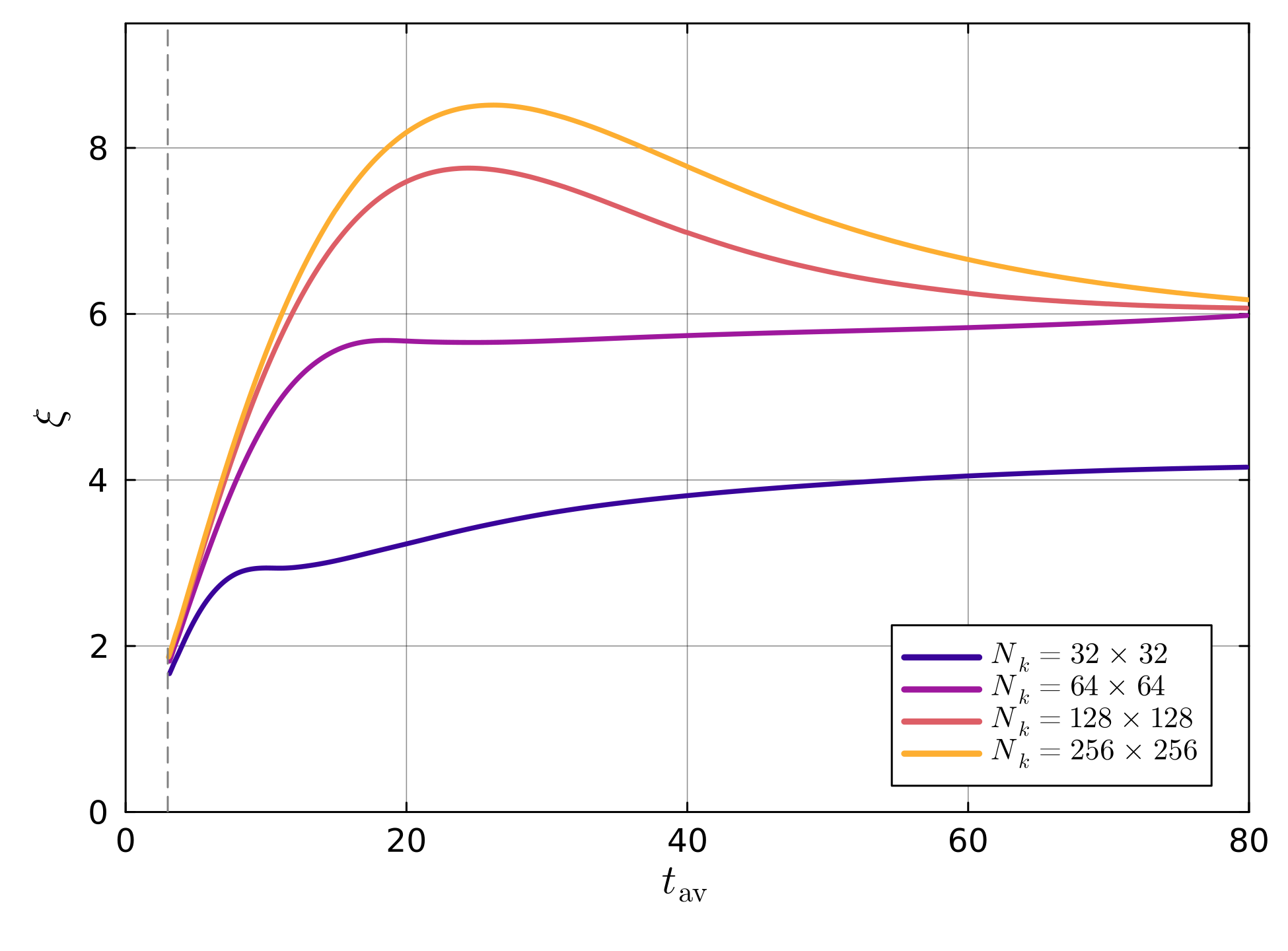}
	\end{minipage}\hfill
	\begin{minipage}[t]{0.49\textwidth}
		\small (b) Electron--phonon model\\[2pt]
		\includegraphics[width=\linewidth]{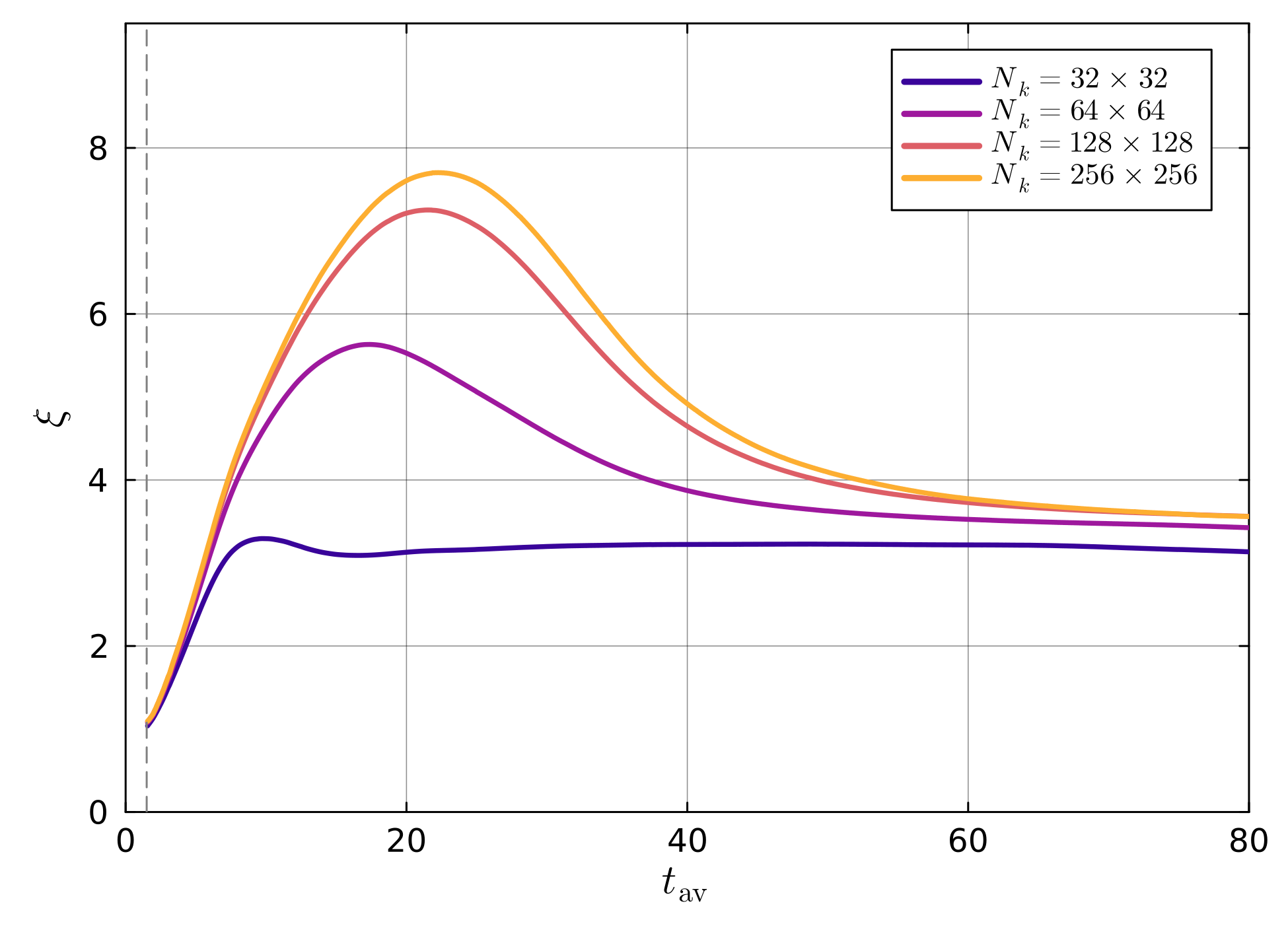}
	\end{minipage}
	\caption{Correlation length $\xi$, Eq.~\eqref{eq:xi_fit}, as a function of $t_{\mathrm{av}}$ for different lattice sizes, (a) of the AFM correlations ($\chi^s$) in the Hubbard model and (b) of the CDW correlations ($\chi^c$) in the electron--phonon model. The vertical dashed lines mark the end of the ramps.
	}
	\label{fig_xi}
\end{figure*}

\paragraph*{Electron--phonon model.}
As in the Hubbard model, some features of the charge response in Fig.~\ref{fig_transient_order}(b) depend only on $t-t'$ and are present for all perturbation times. Close to the diagonal, the same peak and oscillations as in the noninteracting bubble appear, but the oscillations persist much longer than in the Hubbard model. In addition, a second maximum appears at $t-t' \approx 4$: the part of the response mediated by the phonons, which react more slowly than the electrons. In the Hubbard model, the spin response is larger than the noninteracting bubble from the start, already at the sharp peak at $t-t' \approx 0.3$ [Fig.~\ref{fig_finite_size}(a)]. In the electron--phonon model, the charge response follows the noninteracting bubble up to $t-t' \approx 1$ and exceeds it only afterwards [Fig.~\ref{fig_finite_size}(b)], since the phonons, unlike the instantaneous $U$, react only with a delay.

Here as well, a slow component appears, which depends on the perturbation time $t'$. For perturbations during the ramp, the response reaches its maximum around $t \approx 13$, about 12 time units after the end of the ramp, as in the Hubbard model. For later perturbation times, it is weaker and does not show a transient growth. For $t' \gtrsim 20$, the response again depends almost only on $t-t'$. Measured after the same delay $t-t'$, the maximal response to a perturbation during the ramp is more than twice as large as in the stationary regime.

This slow component is the signature of a collective CDW fluctuation at $(\pi,\pi)$, which plays a similar role as the paramagnon: in the nearly thermal state after the ramp, the charge spectrum shows a sharp peak at $(\pi,\pi)$ and $\omega \approx 0$ [Fig.~\ref{fig_spectra}(b)].
The transient enhancement has the same origin as in the Hubbard model. In the static limit, the phonon-mediated interaction is $2V^{0,R}_{\mathbf{q}}(\omega=0) = -4g^2$ for all $\mathbf{q}\neq0$, so that $\chi^c$ obeys the same equation as $\chi^s$ with $U$ replaced by $4g^2$, and the CDW instability occurs at $1+4g^2\chi_{(\pi,\pi)} = 0$. With $g = 0.7$, $4g^2 = 1.96$ is close to $U = 2$, so that both ramps lead to similar
enhancements of the susceptibility.
The phonon coupling reaches its final value after 1.5 time units, while $|\chi_{(\pi,\pi)}|$ again decreases only as the electrons heat up. The transient enhancement thus relies on the same two ingredients as in the Hubbard model: a final state close to the instability and a ramp faster than the heating of the electrons.

While CDW fluctuations couple to the phonons, the slow component of the charge susceptibility decays without oscillating.
The reason is that the $(\pi,\pi)$ phonon is strongly softened close to the instability. In a static estimate, i.e., with $\chi_{(\pi,\pi)}$ taken as frequency independent ($\omega = 0$), the phonon Dyson equation, Eq.~\eqref{eq:phonon_dyson}, with $D^{0,R}_{\mathbf q}(\omega) = 2\Omega_{\mathbf q}/(\omega^2-\Omega_{\mathbf q}^2)$ and $g_{(\pi,\pi)}^2 = \Omega_0 g^2$ [Eq.~\eqref{eq:phonon_dispersion}], gives a dressed propagator of the same form as the bare one, $D^{R}_{(\pi,\pi)}(\omega) = 2\Omega_0/(\omega^2-\Omega_r^2)$, with the renormalized frequency $\Omega_r = \Omega_0\sqrt{1+4g^2\chi_{(\pi,\pi)}}$. It contains the same factor as the denominator of $\chi^c$, which is small close to the instability. The phonon is therefore too slow to complete an oscillation before it decays.

\begin{figure*}[t]
	\begin{minipage}[t]{0.49\textwidth}
		\small (a) Hubbard model\\[2pt]
		\includegraphics[width=\linewidth]{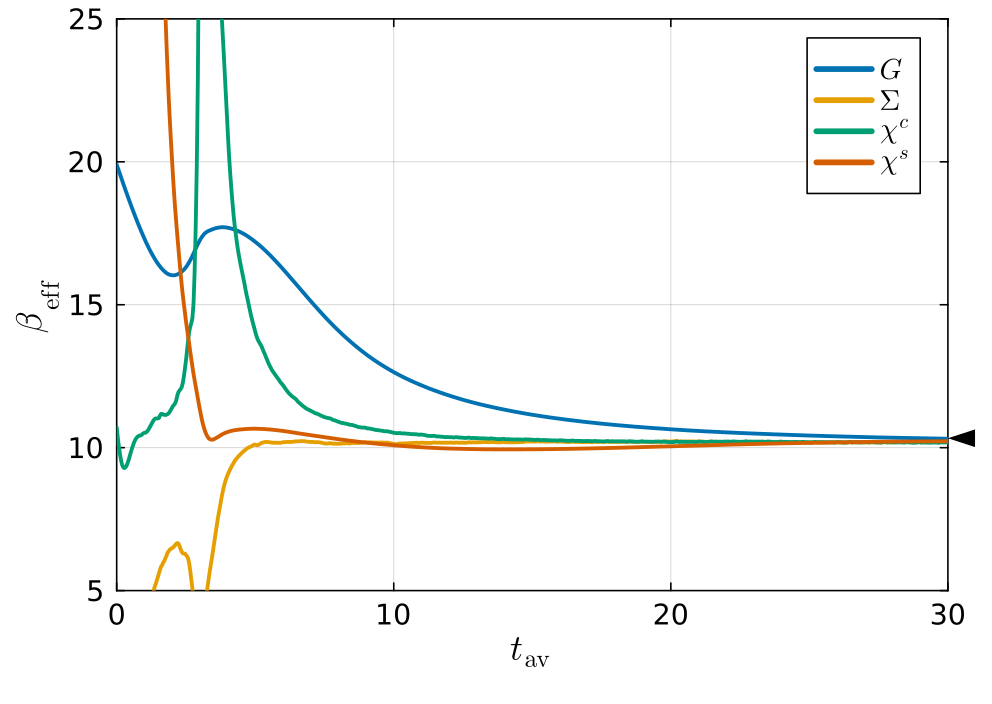}
	\end{minipage}\hfill
	\begin{minipage}[t]{0.49\textwidth}
		\small (b) Electron--phonon model\\[2pt]
		\includegraphics[width=\linewidth]{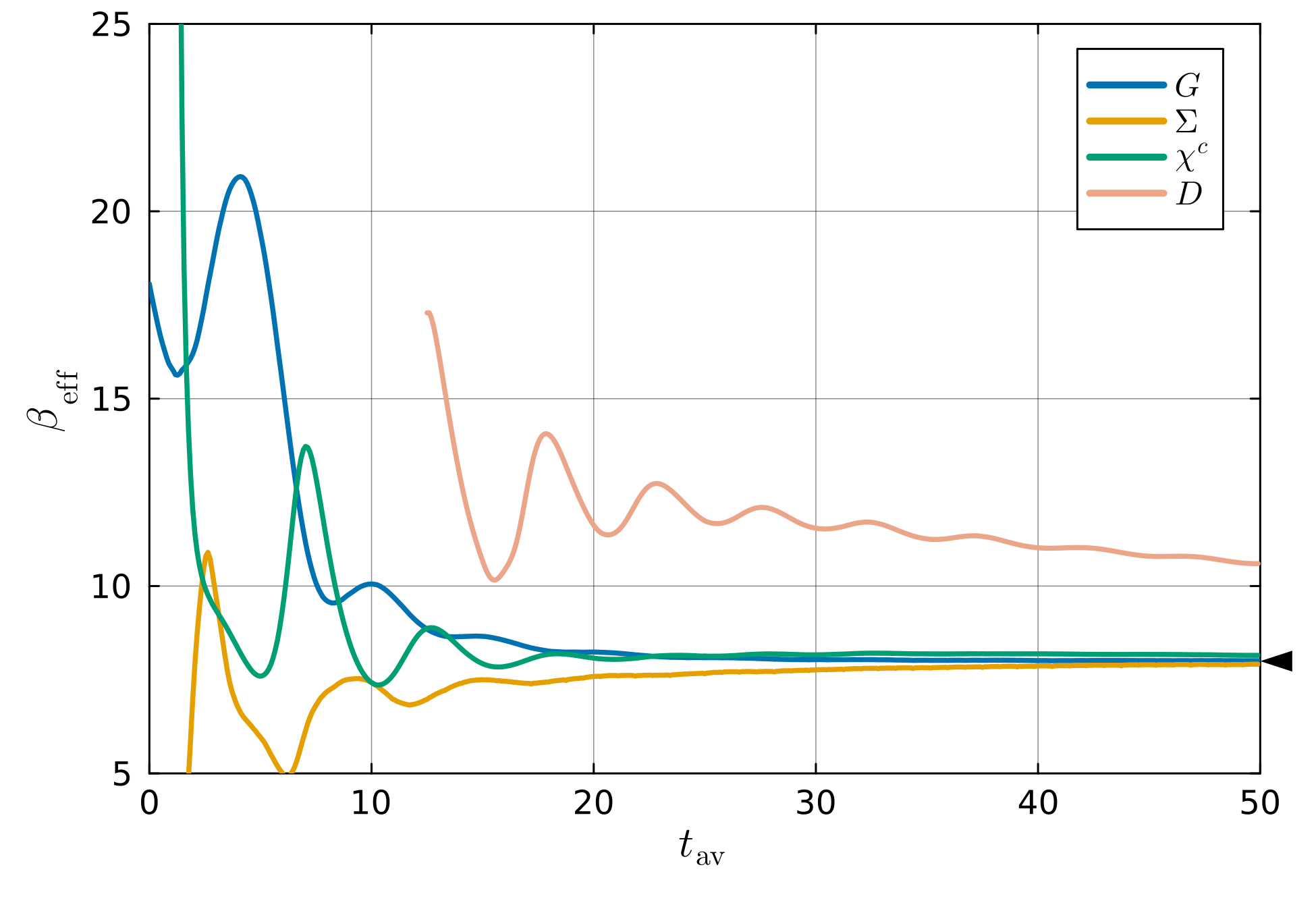}
	\end{minipage}
	\caption{Effective inverse temperature $\beta_{\mathrm{eff}}$ from fits of the local distribution functions in (a) the Hubbard model and (b) the electron--phonon model. The fermionic (bosonic) functions are fitted for $|\omega| \le 5$ ($0.05 \le \omega \le 2$). The initial state has $\beta = 100$. The triangles on the right axis mark $\beta_{\mathrm{eq}}$ obtained from energy matching.}
	\label{fig_temperature}
\end{figure*}

\subsection{Finite-size effects}\label{sec:finite_size}
For the Hubbard model, Fig.~\ref{fig_finite_size}(a) compares cuts of the retarded spin susceptibility at $t' = 2$ for momentum grids from $N_k = 32^2$ to $256^2$. The curves for all grids coincide during the initial peak and separate from each other at later times, where significant finite-size effects become evident. The curve for $32^2$ drops steeply around $t \approx 10$, while the curves for the larger lattices separate gradually.
In the electron--phonon model [Fig.~\ref{fig_finite_size}(b)], the $32^2$ curve separates already from $t \approx 6$ and drops further around $t \approx 10$.
In both models, the results for $N_k=128^2$ and $256^2$ remain close at the maximum of the slow component. The two curves stay closer to each other in the electron--phonon model, which thus converges faster with the lattice size.

Part of this behavior can be understood by looking at the light cone of the correlations. For the Hubbard (electron--phonon) model, Fig.~\ref{fig_cone} shows $\chi^{s,R}_{\mathbf{i},\mathbf{i}+(n,n)}$ ($\chi^{c,R}_{\mathbf{i},\mathbf{i}+(n,n)}$) along the diagonal for the largest lattice.
The response of our weakly correlated systems stays within a light cone, $|n| \le 2(t-t')$, set by the maximal group velocity $|\partial\varepsilon_{\mathbf k}/\partial k_{x}| \le 2t_h$ of the noninteracting electrons. On an $L\times L$ lattice with periodic boundary conditions, the fronts that propagate in opposite directions meet after $t-t' = L/4$, and the response then interferes with its periodic images. For the cuts at $t'=2$ (1.5 in the electron--phonon model), this happens at $t = 10$ (9.5) 
for $L=32$ [dashed lines in Fig.~\ref{fig_finite_size}],
which coincides with the drop of the $32^2$ curves in Fig.~\ref{fig_finite_size}. For $L = 64$, it happens at $t = 18$ (17.5), where the $64^2$ curves show a weaker drop. For $L = 128$ and 256, the light cone reaches the boundary only at $t = 34$ and 66 (33.5 and 65.5), and no prominent feature is visible, presumably because the fronts have decayed by then.
In the electron--phonon model [Fig.~\ref{fig_cone}(b)], the response additionally contains a component confined to the much narrower light cone of the phonons, $|n| \le \Omega_0(t-t')/4$. Since the phonons are much slower than the electrons, the corresponding fronts meet much later, and their interference plays no role here.

In both models, however, the susceptibilities already separate well before the fronts of the light cone meet, an observation which interference effects cannot explain. Before that time, the response itself does not feel the finite system size, but the state that is perturbed does. This is not because the AFM (CDW) correlations are cut off by the system size: in the Hubbard (electron--phonon) model, the correlation length discussed in Sec.~\ref{sec:xi} [Eq.~\eqref{eq:xi_fit}, Fig.~\ref{fig_xi}] reaches at most $\xi \approx 8.5$ ($7.7$) lattice spacings within the simulated times, much smaller than $L$. Instead, the electrons themselves feel the grid: the initial state at $\beta = 100$ is so cold that the momentum grid does not fully resolve the Fermi surface. Already for noninteracting electrons at this temperature, the static bubble at $(\pi,\pi)$ deviates by $3.6\%$ ($0.3\%$) on the $128^2$ ($256^2$) grid from its value on a $2048^2$ grid, whereas at $\beta = 10$ all grids agree. $G$, $\Sigma$, and the bubble therefore depend slightly on $N_k$. Close to the instability, this small change of the bubble is strongly amplified, by the factor $1/(1+U\chi)$ in $\chi^s = \chi/(1+U\chi)$, and by $1/(1+4g^2\chi)$ for $\chi^c$ in the static limit. For example, in the Hubbard model at $t = 35$, the static bubble differs by only $0.7\%$ between $N_k = 128^2$ and $256^2$, whereas $1/(1+U\chi)$ increases from 28 to 35.

\subsection{Correlation length}\label{sec:xi}
The AFM correlations in the Hubbard model, measured by the equal-time spin correlation $\ii\chi^{s,<}_{(\pi,\pi)}(t,t) \simeq \tfrac12\sum_{\mathbf j}(-1)^{j_x+j_y}\langle S^z_{\mathbf 0}(t)\,S^z_{\mathbf j}(t)\rangle$, also grow after the ramp and then decay toward their equilibrium value. Figure~\ref{fig_xi} shows the corresponding correlation length, defined in Eq.~\eqref{eq:xi_fit}. On the largest grid, $N_k = 256^2$, $\xi$ grows rapidly after the ramp, reaches a maximum of about 8.5 lattice spacings at $t \approx 26$, and then decreases slowly, to about 6.2 at $t = 80$. Since $\xi$ stays finite and much smaller than $L$ at all times, no long-range order develops: the system shows only a transient enhancement of short-range AFM order. We also note that both the equal-time correlation and $\xi$ peak later than the response in Fig.~\ref{fig_transient_order}(a). Roughly speaking, one can view the spin fluctuations at successive times $t'$ as a succession of perturbations, each followed by a response given by the cut along $t$ in Fig.~\ref{fig_transient_order}(a). The spatial correlations at time $t$ add up these responses. They grow as long as the enhanced, slow responses to the early perturbations accumulate, and decrease once these have decayed, since later perturbations trigger only the weaker response of the stationary regime.

The CDW correlations in the electron--phonon model also show transiently enhanced short-range order: their correlation length grows as in the Hubbard model, peaks at $\xi \approx 7.7$ around $t \approx 22$, and then drops faster, to about 4.1 at $t = 50$ and about 3.6 at $t = 80$ [Fig.~\ref{fig_xi}(b)].
As for the cuts in Fig.~\ref{fig_finite_size}, the curves for $N_k = 128^2$ and $256^2$ are closer to each other in the electron--phonon model [Fig.~\ref{fig_xi}(b)] than in the Hubbard model [Fig.~\ref{fig_xi}(a)].

\subsection{Effective temperature}\label{sec:temperature}

Figure~\ref{fig_temperature} shows the effective inverse temperatures $\beta_{\mathrm{eff}}$ obtained from fits of the local (momentum-summed) distribution functions, Eq.~\eqref{eq:distribution}, as a function of $t_{\mathrm{av}}$. Three caveats apply to these fits. First, if the ramp starts at $t = 0$, the window in $t_{\mathrm{rel}}$ available for the Fourier transform is short at early times. A sharp cut at the window edge produces ringing in the spectra, and a short window limits the frequency resolution. We therefore ran additional simulations in which the ramp is delayed, all for $N_k = 128\times128$. The system first remains in its initial equilibrium state for 100 time units in the electron--phonon model and for 40 time units in the Hubbard model. Then the ramp is applied and the induced dynamics simulated. In this setup, longer relative times become available. In the electron--phonon model, a longer waiting time before the ramp is needed because the two-time functions decay more slowly in $t_{\mathrm{rel}}$, compared to the Hubbard model. Second, a two-time function $X(t,t') \propto \langle A(t)\,B(t')\rangle$ describes correlations of the system between two times, which can be far apart and correspond to very different temperatures, for example before and after the ramp. If the function decays slowly in $t_{\mathrm{rel}}$, its fitted $\beta_{\mathrm{eff}}$ is therefore not a well-defined temperature. This is a problem mainly at early times, when the state changes rapidly and the cold state before the ramp is still within the range of the relative times. Third, far from equilibrium, the distribution functions typically do not exhibit a Fermi--Dirac or Bose--Einstein form, in which case the fit parameter $\beta_{\mathrm{eff}}$ cannot be interpreted as an effective inverse temperature. For these reasons, the effective temperatures at early times should be interpreted with caution.

We first consider the Hubbard model [Fig.~\ref{fig_temperature}(a)]. At early times, the fitted temperatures vary strongly, and $\chi^s$ and $\chi^c$ show sharp peaks at the start and at the end of the ramp. As explained above, the system is then far from equilibrium, and these values should not be regarded as well-defined inverse temperatures. Once the ramp is over, the effective temperatures for the different functions approach the same value, but on different time scales. $\Sigma$ shows the final temperature within a few inverse hoppings after the end of the ramp, $\chi^c$ within about 20 inverse hoppings, and $G$ within 30 inverse hoppings. $\chi^s$ slightly overheats: its $\beta_{\mathrm{eff}}$ drops below the final value of 10.2, to about 9.9 at $t_{\mathrm{av}} \approx 14$. The common final value of the four effective temperatures agrees with the temperature of the thermalized system obtained from the energy matching condition. The effective temperature of $G$ is determined by the electrons close to the Fermi surface, which are still colder than the final temperature when $\Sigma$ and $\chi^c$ have already reached it. Because of nesting, these electrons contribute strongly to the bubble at $(\pi,\pi)$, so that the response remains enhanced as long as they are cold (Sec.~\ref{sec:transient}).

We now turn to the electron--phonon model [Fig.~\ref{fig_temperature}(b)]. At early times, the fitted temperatures again vary strongly. They also oscillate for some time after the ramp, most visibly for $D$. These oscillations are not ringing associated with the window of the Fourier transform, since the window is long at these times. All four functions oscillate with the same period ($\approx 5$). After a fast change of the coupling, momentum-averaged quantities are expected to oscillate at twice the renormalized phonon frequency~\cite{Murakami2015,Abdurazakov2018,Sroda2025}. The observed period is longer than the bare value $2\pi/(2\Omega_0) \approx 3.1$ at $(\pi,\pi)$, consistent with a softening of the phonons. Once the ramp is over, the effective temperature extracted from the electronic subsystem settles on a time scale similar to that in the Hubbard model. $G$ reaches the temperature obtained from energy matching around $t_{\mathrm{av}} = 30$, and the effective temperatures of $\Sigma$ and $\chi^c$ lie close to it. The phonon propagator behaves differently. It stays colder than the thermalized state and heats up only slowly, without reaching the final temperature within the simulated time window. There are two probable causes. The first is the fit itself: $D$ decays more slowly in $t_{\mathrm{rel}}$ than the other propagators, so that its distribution function mixes in the colder states of earlier times for longer. The second is a bottleneck in the thermalization of the low-energy phonons, which was analyzed in detail for a coupling quench in Ref.~\cite{Sroda2026}. Since the fit to the Bose function is determined mainly by the lowest frequencies, where the distribution function is largest, these nonthermal cold modes make $D$ appear colder.

In both models, the electrons thus reach a temperature consistent with energy conservation within 30 to 40 inverse hopping times. This is comparable to the perturbation time beyond which the spin and charge responses depend only on $t-t'$ ($t' \gtrsim 20$, Sec.~\ref{sec:transient}). The low-energy, long-wavelength phonons and the correlation lengths [Fig.~\ref{fig_xi}], however, relax more slowly.

\section{Conclusions}\label{conclusions}
In this work, we studied how the spin and charge susceptibilities evolve in time in two half-filled models on the square lattice, the Hubbard model and an electron--phonon model with acoustic phonons. In both, a ramp of the interaction or of the electron--phonon coupling brings the system close to an ordering instability, antiferromagnetic in the first case and charge-density-wave in the second. We solved the fluctuation-exchange (FLEX) and Migdal approximations self-consistently, with nonequilibrium Green's functions compressed as quantics tensor trains. This allowed us to compute the two-time susceptibilities with full momentum resolution on lattices of up to $256\times256$ sites.

Both models show a transient enhancement of the response at $(\pi,\pi)$. This phenomenon requires two ingredients: a final state close to the instability, and a ramp faster than the heating of the electrons, so that the interaction or the coupling reaches its final value while the electrons are still cold. In the Hubbard model, the static denominator $1+U\chi_{(\pi,\pi)}$ of the Dyson equation for $\chi^s$, Eq.~\eqref{eq:bosonic_dyson}, is then smaller than in the thermalized state, and even slightly negative for about 15 hopping times. As a result, the retarded spin susceptibility $\chi^{s,R}_{(\pi,\pi)}(t,t')$ for a perturbation during the ramp reaches a maximum about twice as large as in the stationary regime, and also the correlation length $\xi$ is transiently enhanced. This behavior corresponds to the transient short-range order discussed in Ref.~\cite{Bauer2015}. In FLEX, the spin fluctuations feed back on the self-energy, in contrast to the local second-order treatment of Ref.~\cite{Bauer2015}, and this likely explains why instead of an instability we only find a modest transient enhancement of correlations for the chosen ramp setup. In both models, the slow dynamics originates from a soft collective mode. This mode is a paramagnon in the Hubbard model and a collective CDW fluctuation in the electron--phonon model.

The effective temperatures give a complementary view of the thermalization process. In the Hubbard model, the self-energy very quickly reaches the final temperature after the end of the ramp, while the Green's function takes about 30 to 40 time units. Their common final value agrees with the temperature expected from energy conservation. In the electron--phonon model, the quantities related to the electronic subsystem essentially thermalize on the same time scale, but the effective temperature extracted from the phonon propagator remains colder and continues to increase at the longest simulation times, which points to a slow relaxation of the low-energy phonons. Thermalization is not complete as long as some effective temperatures deviate from the thermal value. Also, it is noteworthy that for both models, the correlation lengths keep evolving even after the electronic distribution functions have
essentially settled.

Our results also revealed a practical challenge. Close to the instability, the susceptibilities depend strongly on the lattice size. This is not only a light-cone effect: the size dependence appears before the nonlocal correlations reach their periodic images. The reason is that the cold initial state makes the particle--hole bubble weakly size dependent, and the small denominator in the susceptibilities amplifies this dependence. A strong transient enhancement requires a cold initial state, and quantitative results in this regime require momentum grids that are fine enough to resolve the Fermi surface of this state. Large lattices, and hence compressed representations of the Green's functions, are therefore essential for accurate simulations.

This study can be extended in several directions. First, the susceptibilities are of random-phase-approximation type, and the missing vertex corrections likely lead to an overestimation of the enhancement, so that it would be interesting to go beyond this approximation. Second, larger systems could be reached by compressing the momentum dependence in addition to the time dependence. Third, an implementation which allows for symmetry breaking would make it possible to enter the charge-density-wave phase of the electron--phonon model or the $d$-wave superconducting phase of the weakly coupled doped Hubbard model, where the interplay between superconductivity and antiferromagnetic spin fluctuations could be studied in nonequilibrium settings.

\begin{acknowledgments}
This work was supported by SNSF Grant No.\ 2000-1-240023 and by the German Research Unit QUAST (FOR 5249) via SNSF Grant No.\ 2000-0E\_238272. 
The calculations were performed on the beo05 cluster at the University of Fribourg.
We used Claude Fable 5.1 and 5.2, Claude Opus 5.4 and 5.5, and Grok 4.6 and 4.7 for scientific reasoning and interpretation, writing parts of the simulation and analysis code, the generation of plotting scripts, drafting and revising scientific claims, literature synthesis, and proofreading and editing. The artificial intelligence tools were directed through iterative prompting and all generated content was reviewed, edited, and verified by the authors.
\end{acknowledgments}

\bibliography{ref}

%apsrev4-2.bst 2019-01-14 (MD) hand-edited version of apsrev4-1.bst
%Control: key (0)
%Control: author (8) initials jnrlst
%Control: editor formatted (1) identically to author
%Control: production of article title (0) allowed
%Control: page (0) single
%Control: year (1) truncated
%Control: production of eprint (0) enabled
\begin{thebibliography}{64}%
\makeatletter
\providecommand \@ifxundefined [1]{%
 \@ifx{#1\undefined}
}%
\providecommand \@ifnum [1]{%
 \ifnum #1\expandafter \@firstoftwo
 \else \expandafter \@secondoftwo
 \fi
}%
\providecommand \@ifx [1]{%
 \ifx #1\expandafter \@firstoftwo
 \else \expandafter \@secondoftwo
 \fi
}%
\providecommand \natexlab [1]{#1}%
\providecommand \enquote  [1]{``#1''}%
\providecommand \bibnamefont  [1]{#1}%
\providecommand \bibfnamefont [1]{#1}%
\providecommand \citenamefont [1]{#1}%
\providecommand \href@noop [0]{\@secondoftwo}%
\providecommand \href [0]{\begingroup \@sanitize@url \@href}%
\providecommand \@href[1]{\@@startlink{#1}\@@href}%
\providecommand \@@href[1]{\endgroup#1\@@endlink}%
\providecommand \@sanitize@url [0]{\catcode `\\12\catcode `\$12\catcode
  `\&12\catcode `\#12\catcode `\^12\catcode `\_12\catcode `\%12\relax}%
\providecommand \@@startlink[1]{}%
\providecommand \@@endlink[0]{}%
\providecommand \url  [0]{\begingroup\@sanitize@url \@url }%
\providecommand \@url [1]{\endgroup\@href {#1}{\urlprefix }}%
\providecommand \urlprefix  [0]{URL }%
\providecommand \Eprint [0]{\href }%
\providecommand \doibase [0]{https://doi.org/}%
\providecommand \selectlanguage [0]{\@gobble}%
\providecommand \bibinfo  [0]{\@secondoftwo}%
\providecommand \bibfield  [0]{\@secondoftwo}%
\providecommand \translation [1]{[#1]}%
\providecommand \BibitemOpen [0]{}%
\providecommand \bibitemStop [0]{}%
\providecommand \bibitemNoStop [0]{.\EOS\space}%
\providecommand \EOS [0]{\spacefactor3000\relax}%
\providecommand \BibitemShut  [1]{\csname bibitem#1\endcsname}%
\let\auto@bib@innerbib\@empty
%</preamble>
\bibitem [{\citenamefont {Berges}\ \emph {et~al.}(2004)\citenamefont {Berges},
  \citenamefont {Bors{\'a}nyi},\ and\ \citenamefont {Wetterich}}]{Berges2004}%
  \BibitemOpen
  \bibfield  {author} {\bibinfo {author} {\bibfnamefont {J.}~\bibnamefont
  {Berges}}, \bibinfo {author} {\bibfnamefont {S.}~\bibnamefont
  {Bors{\'a}nyi}},\ and\ \bibinfo {author} {\bibfnamefont {C.}~\bibnamefont
  {Wetterich}},\ }\bibfield  {title} {\bibinfo {title} {Prethermalization},\
  }\href {https://doi.org/10.1103/PhysRevLett.93.142002} {\bibfield  {journal}
  {\bibinfo  {journal} {Phys. Rev. Lett.}\ }\textbf {\bibinfo {volume} {93}},\
  \bibinfo {pages} {142002} (\bibinfo {year} {2004})}\BibitemShut {NoStop}%
\bibitem [{\citenamefont {Moeckel}\ and\ \citenamefont
  {Kehrein}(2008)}]{Moeckel2008}%
  \BibitemOpen
  \bibfield  {author} {\bibinfo {author} {\bibfnamefont {M.}~\bibnamefont
  {Moeckel}}\ and\ \bibinfo {author} {\bibfnamefont {S.}~\bibnamefont
  {Kehrein}},\ }\bibfield  {title} {\bibinfo {title} {Interaction quench in the
  hubbard model},\ }\href {https://doi.org/10.1103/PhysRevLett.100.175702}
  {\bibfield  {journal} {\bibinfo  {journal} {Phys. Rev. Lett.}\ }\textbf
  {\bibinfo {volume} {100}},\ \bibinfo {pages} {175702} (\bibinfo {year}
  {2008})}\BibitemShut {NoStop}%
\bibitem [{\citenamefont {{\'S}roda}\ \emph {et~al.}(2025)\citenamefont
  {{\'S}roda}, \citenamefont {Inayoshi}, \citenamefont {Shinaoka},\ and\
  \citenamefont {Werner}}]{Sroda2025}%
  \BibitemOpen
  \bibfield  {author} {\bibinfo {author} {\bibfnamefont {M.}~\bibnamefont
  {{\'S}roda}}, \bibinfo {author} {\bibfnamefont {K.}~\bibnamefont {Inayoshi}},
  \bibinfo {author} {\bibfnamefont {H.}~\bibnamefont {Shinaoka}},\ and\
  \bibinfo {author} {\bibfnamefont {P.}~\bibnamefont {Werner}},\ }\bibfield
  {title} {\bibinfo {title} {Memory-efficient nonequilibrium {G}reen's function
  framework built on quantics tensor trains},\ }\href
  {https://doi.org/10.1103/dxfb-b3l5} {\bibfield  {journal} {\bibinfo
  {journal} {Phys. Rev. Lett.}\ }\textbf {\bibinfo {volume} {135}},\ \bibinfo
  {pages} {226501} (\bibinfo {year} {2025})},\ \Eprint
  {https://arxiv.org/abs/2412.14032} {arXiv:2412.14032 [cond-mat.str-el]}
  \BibitemShut {NoStop}%
\bibitem [{\citenamefont {Sensarma}\ \emph {et~al.}(2010)\citenamefont
  {Sensarma}, \citenamefont {Pekker}, \citenamefont {Altman}, \citenamefont
  {Demler}, \citenamefont {Strohmaier}, \citenamefont {Greif}, \citenamefont
  {J{\"o}rdens}, \citenamefont {Tarruell}, \citenamefont {Moritz},\ and\
  \citenamefont {Esslinger}}]{Sensarma2010}%
  \BibitemOpen
  \bibfield  {author} {\bibinfo {author} {\bibfnamefont {R.}~\bibnamefont
  {Sensarma}}, \bibinfo {author} {\bibfnamefont {D.}~\bibnamefont {Pekker}},
  \bibinfo {author} {\bibfnamefont {E.}~\bibnamefont {Altman}}, \bibinfo
  {author} {\bibfnamefont {E.}~\bibnamefont {Demler}}, \bibinfo {author}
  {\bibfnamefont {N.}~\bibnamefont {Strohmaier}}, \bibinfo {author}
  {\bibfnamefont {D.}~\bibnamefont {Greif}}, \bibinfo {author} {\bibfnamefont
  {R.}~\bibnamefont {J{\"o}rdens}}, \bibinfo {author} {\bibfnamefont
  {L.}~\bibnamefont {Tarruell}}, \bibinfo {author} {\bibfnamefont
  {H.}~\bibnamefont {Moritz}},\ and\ \bibinfo {author} {\bibfnamefont
  {T.}~\bibnamefont {Esslinger}},\ }\bibfield  {title} {\bibinfo {title}
  {Lifetime of double occupancies in the fermi-hubbard model},\ }\href
  {https://doi.org/10.1103/PhysRevB.82.224302} {\bibfield  {journal} {\bibinfo
  {journal} {Phys. Rev. B}\ }\textbf {\bibinfo {volume} {82}},\ \bibinfo
  {pages} {224302} (\bibinfo {year} {2010})}\BibitemShut {NoStop}%
\bibitem [{\citenamefont {Eckstein}\ and\ \citenamefont
  {Werner}(2011)}]{Eckstein2011}%
  \BibitemOpen
  \bibfield  {author} {\bibinfo {author} {\bibfnamefont {M.}~\bibnamefont
  {Eckstein}}\ and\ \bibinfo {author} {\bibfnamefont {P.}~\bibnamefont
  {Werner}},\ }\bibfield  {title} {\bibinfo {title} {Thermalization of a
  pump-excited mott insulator},\ }\href
  {https://doi.org/10.1103/PhysRevB.84.035122} {\bibfield  {journal} {\bibinfo
  {journal} {Phys. Rev. B}\ }\textbf {\bibinfo {volume} {84}},\ \bibinfo
  {pages} {035122} (\bibinfo {year} {2011})}\BibitemShut {NoStop}%
\bibitem [{\citenamefont {Sentef}\ \emph {et~al.}(2013)\citenamefont {Sentef},
  \citenamefont {Kemper}, \citenamefont {Moritz}, \citenamefont {Freericks},
  \citenamefont {Shen},\ and\ \citenamefont {Devereaux}}]{Sentef2013}%
  \BibitemOpen
  \bibfield  {author} {\bibinfo {author} {\bibfnamefont {M.}~\bibnamefont
  {Sentef}}, \bibinfo {author} {\bibfnamefont {A.~F.}\ \bibnamefont {Kemper}},
  \bibinfo {author} {\bibfnamefont {B.}~\bibnamefont {Moritz}}, \bibinfo
  {author} {\bibfnamefont {J.~K.}\ \bibnamefont {Freericks}}, \bibinfo {author}
  {\bibfnamefont {Z.-X.}\ \bibnamefont {Shen}},\ and\ \bibinfo {author}
  {\bibfnamefont {T.~P.}\ \bibnamefont {Devereaux}},\ }\bibfield  {title}
  {\bibinfo {title} {Examining electron-boson coupling using time-resolved
  spectroscopy},\ }\href {https://doi.org/10.1103/PhysRevX.3.041033} {\bibfield
   {journal} {\bibinfo  {journal} {Phys. Rev. X}\ }\textbf {\bibinfo {volume}
  {3}},\ \bibinfo {pages} {041033} (\bibinfo {year} {2013})}\BibitemShut
  {NoStop}%
\bibitem [{\citenamefont {Murakami}\ \emph {et~al.}(2015)\citenamefont
  {Murakami}, \citenamefont {Werner}, \citenamefont {Tsuji},\ and\
  \citenamefont {Aoki}}]{Murakami2015}%
  \BibitemOpen
  \bibfield  {author} {\bibinfo {author} {\bibfnamefont {Y.}~\bibnamefont
  {Murakami}}, \bibinfo {author} {\bibfnamefont {P.}~\bibnamefont {Werner}},
  \bibinfo {author} {\bibfnamefont {N.}~\bibnamefont {Tsuji}},\ and\ \bibinfo
  {author} {\bibfnamefont {H.}~\bibnamefont {Aoki}},\ }\bibfield  {title}
  {\bibinfo {title} {Interaction quench in the holstein model: Thermalization
  crossover from electron- to phonon-dominated relaxation},\ }\href
  {https://doi.org/10.1103/PhysRevB.91.045128} {\bibfield  {journal} {\bibinfo
  {journal} {Phys. Rev. B}\ }\textbf {\bibinfo {volume} {91}},\ \bibinfo
  {pages} {045128} (\bibinfo {year} {2015})}\BibitemShut {NoStop}%
\bibitem [{\citenamefont {Avigo}\ \emph {et~al.}(2016)\citenamefont {Avigo},
  \citenamefont {Thirupathaiah}, \citenamefont {Ligges}, \citenamefont {Wolf},
  \citenamefont {Fink},\ and\ \citenamefont {Bovensiepen}}]{Avigo2016}%
  \BibitemOpen
  \bibfield  {author} {\bibinfo {author} {\bibfnamefont {I.}~\bibnamefont
  {Avigo}}, \bibinfo {author} {\bibfnamefont {S.}~\bibnamefont
  {Thirupathaiah}}, \bibinfo {author} {\bibfnamefont {M.}~\bibnamefont
  {Ligges}}, \bibinfo {author} {\bibfnamefont {T.}~\bibnamefont {Wolf}},
  \bibinfo {author} {\bibfnamefont {J.}~\bibnamefont {Fink}},\ and\ \bibinfo
  {author} {\bibfnamefont {U.}~\bibnamefont {Bovensiepen}},\ }\bibfield
  {title} {\bibinfo {title} {Doping dependence and electron-boson coupling in
  the ultrafast relaxation of hot electron populations in ba(fe1-xcox)2as2},\
  }\href {https://doi.org/10.1088/1367-2630/18/9/093028} {\bibfield  {journal}
  {\bibinfo  {journal} {New Journal of Physics}\ }\textbf {\bibinfo {volume}
  {18}},\ \bibinfo {pages} {093028} (\bibinfo {year} {2016})}\BibitemShut
  {NoStop}%
\bibitem [{\citenamefont {\'Sroda}\ and\ \citenamefont
  {Werner}(2026)}]{Sroda2026}%
  \BibitemOpen
  \bibfield  {author} {\bibinfo {author} {\bibfnamefont {M.}~\bibnamefont
  {\'Sroda}}\ and\ \bibinfo {author} {\bibfnamefont {P.}~\bibnamefont
  {Werner}},\ }\href@noop {} {\bibinfo {title} {Nonequilibrium electron-phonon
  dynamics with high momentum resolution: Thermalization bottlenecks and the
  effects of phonon dispersion}} (\bibinfo {year} {2026}),\ \Eprint
  {https://arxiv.org/abs/2606.28855} {arXiv:2606.28855 [cond-mat.str-el]}
  \BibitemShut {NoStop}%
\bibitem [{\citenamefont {Eckstein}\ \emph {et~al.}(2009)\citenamefont
  {Eckstein}, \citenamefont {Kollar},\ and\ \citenamefont
  {Werner}}]{Eckstein2009}%
  \BibitemOpen
  \bibfield  {author} {\bibinfo {author} {\bibfnamefont {M.}~\bibnamefont
  {Eckstein}}, \bibinfo {author} {\bibfnamefont {M.}~\bibnamefont {Kollar}},\
  and\ \bibinfo {author} {\bibfnamefont {P.}~\bibnamefont {Werner}},\
  }\bibfield  {title} {\bibinfo {title} {Thermalization after an interaction
  quench in the hubbard model},\ }\href
  {https://doi.org/10.1103/PhysRevLett.103.056403} {\bibfield  {journal}
  {\bibinfo  {journal} {Phys. Rev. Lett.}\ }\textbf {\bibinfo {volume} {103}},\
  \bibinfo {pages} {056403} (\bibinfo {year} {2009})}\BibitemShut {NoStop}%
\bibitem [{\citenamefont {Schir{\'o}}\ and\ \citenamefont
  {Fabrizio}(2011)}]{Schiro2011}%
  \BibitemOpen
  \bibfield  {author} {\bibinfo {author} {\bibfnamefont {M.}~\bibnamefont
  {Schir{\'o}}}\ and\ \bibinfo {author} {\bibfnamefont {M.}~\bibnamefont
  {Fabrizio}},\ }\bibfield  {title} {\bibinfo {title} {Quantum quenches in the
  hubbard model: Time-dependent mean-field theory and the role of quantum
  fluctuations},\ }\href {https://doi.org/10.1103/PhysRevB.83.165105}
  {\bibfield  {journal} {\bibinfo  {journal} {Phys. Rev. B}\ }\textbf {\bibinfo
  {volume} {83}},\ \bibinfo {pages} {165105} (\bibinfo {year}
  {2011})}\BibitemShut {NoStop}%
\bibitem [{\citenamefont {Heyl}(2014)}]{Heyl2014}%
  \BibitemOpen
  \bibfield  {author} {\bibinfo {author} {\bibfnamefont {M.}~\bibnamefont
  {Heyl}},\ }\bibfield  {title} {\bibinfo {title} {Dynamical quantum phase
  transitions in systems with broken-symmetry phases},\ }\href
  {https://doi.org/10.1103/PhysRevLett.113.205701} {\bibfield  {journal}
  {\bibinfo  {journal} {Phys. Rev. Lett.}\ }\textbf {\bibinfo {volume} {113}},\
  \bibinfo {pages} {205701} (\bibinfo {year} {2014})}\BibitemShut {NoStop}%
\bibitem [{\citenamefont {Canovi}\ \emph {et~al.}(2014)\citenamefont {Canovi},
  \citenamefont {Werner},\ and\ \citenamefont {Eckstein}}]{Canovi2014}%
  \BibitemOpen
  \bibfield  {author} {\bibinfo {author} {\bibfnamefont {E.}~\bibnamefont
  {Canovi}}, \bibinfo {author} {\bibfnamefont {P.}~\bibnamefont {Werner}},\
  and\ \bibinfo {author} {\bibfnamefont {M.}~\bibnamefont {Eckstein}},\
  }\bibfield  {title} {\bibinfo {title} {First-order dynamical phase
  transitions},\ }\href {https://doi.org/10.1103/PhysRevLett.113.265702}
  {\bibfield  {journal} {\bibinfo  {journal} {Phys. Rev. Lett.}\ }\textbf
  {\bibinfo {volume} {113}},\ \bibinfo {pages} {265702} (\bibinfo {year}
  {2014})}\BibitemShut {NoStop}%
\bibitem [{\citenamefont {Berges}\ \emph {et~al.}(2008)\citenamefont {Berges},
  \citenamefont {Rothkopf},\ and\ \citenamefont {Schmidt}}]{Berges2008}%
  \BibitemOpen
  \bibfield  {author} {\bibinfo {author} {\bibfnamefont {J.}~\bibnamefont
  {Berges}}, \bibinfo {author} {\bibfnamefont {A.}~\bibnamefont {Rothkopf}},\
  and\ \bibinfo {author} {\bibfnamefont {J.}~\bibnamefont {Schmidt}},\
  }\bibfield  {title} {\bibinfo {title} {Nonthermal fixed points: Effective
  weak coupling for strongly correlated systems far from equilibrium},\ }\href
  {https://doi.org/10.1103/PhysRevLett.101.041603} {\bibfield  {journal}
  {\bibinfo  {journal} {Phys. Rev. Lett.}\ }\textbf {\bibinfo {volume} {101}},\
  \bibinfo {pages} {041603} (\bibinfo {year} {2008})}\BibitemShut {NoStop}%
\bibitem [{\citenamefont {Tsuji}\ \emph {et~al.}(2013)\citenamefont {Tsuji},
  \citenamefont {Eckstein},\ and\ \citenamefont {Werner}}]{Tsuji2013}%
  \BibitemOpen
  \bibfield  {author} {\bibinfo {author} {\bibfnamefont {N.}~\bibnamefont
  {Tsuji}}, \bibinfo {author} {\bibfnamefont {M.}~\bibnamefont {Eckstein}},\
  and\ \bibinfo {author} {\bibfnamefont {P.}~\bibnamefont {Werner}},\
  }\bibfield  {title} {\bibinfo {title} {Nonthermal antiferromagnetic order and
  nonequilibrium criticality in the hubbard model},\ }\href
  {https://doi.org/10.1103/PhysRevLett.110.136404} {\bibfield  {journal}
  {\bibinfo  {journal} {Phys. Rev. Lett.}\ }\textbf {\bibinfo {volume} {110}},\
  \bibinfo {pages} {136404} (\bibinfo {year} {2013})}\BibitemShut {NoStop}%
\bibitem [{\citenamefont {Werner}\ \emph {et~al.}(2012)\citenamefont {Werner},
  \citenamefont {Tsuji},\ and\ \citenamefont {Eckstein}}]{Werner2012}%
  \BibitemOpen
  \bibfield  {author} {\bibinfo {author} {\bibfnamefont {P.}~\bibnamefont
  {Werner}}, \bibinfo {author} {\bibfnamefont {N.}~\bibnamefont {Tsuji}},\ and\
  \bibinfo {author} {\bibfnamefont {M.}~\bibnamefont {Eckstein}},\ }\bibfield
  {title} {\bibinfo {title} {Nonthermal symmetry-broken states in the strongly
  interacting hubbard model},\ }\href
  {https://doi.org/10.1103/PhysRevB.86.205101} {\bibfield  {journal} {\bibinfo
  {journal} {Phys. Rev. B}\ }\textbf {\bibinfo {volume} {86}},\ \bibinfo
  {pages} {205101} (\bibinfo {year} {2012})}\BibitemShut {NoStop}%
\bibitem [{\citenamefont {Li}\ \emph {et~al.}(2018)\citenamefont {Li},
  \citenamefont {Strand}, \citenamefont {Werner},\ and\ \citenamefont
  {Eckstein}}]{Li2018}%
  \BibitemOpen
  \bibfield  {author} {\bibinfo {author} {\bibfnamefont {J.}~\bibnamefont
  {Li}}, \bibinfo {author} {\bibfnamefont {H.}~\bibnamefont {Strand}}, \bibinfo
  {author} {\bibfnamefont {P.}~\bibnamefont {Werner}},\ and\ \bibinfo {author}
  {\bibfnamefont {M.}~\bibnamefont {Eckstein}},\ }\bibfield  {title} {\bibinfo
  {title} {Theory of photoinduced ultrafast switching to a spin-orbital ordered
  hidden phase},\ }\href {https://doi.org/10.1038/s41467-018-07051-x}
  {\bibfield  {journal} {\bibinfo  {journal} {Nature Communications}\ }\textbf
  {\bibinfo {volume} {9}},\ \bibinfo {pages} {4581} (\bibinfo {year}
  {2018})}\BibitemShut {NoStop}%
\bibitem [{\citenamefont {Kaneko}\ \emph {et~al.}(2019)\citenamefont {Kaneko},
  \citenamefont {Shirakawa}, \citenamefont {Sorella},\ and\ \citenamefont
  {Yunoki}}]{Kaneko2019}%
  \BibitemOpen
  \bibfield  {author} {\bibinfo {author} {\bibfnamefont {T.}~\bibnamefont
  {Kaneko}}, \bibinfo {author} {\bibfnamefont {T.}~\bibnamefont {Shirakawa}},
  \bibinfo {author} {\bibfnamefont {S.}~\bibnamefont {Sorella}},\ and\ \bibinfo
  {author} {\bibfnamefont {S.}~\bibnamefont {Yunoki}},\ }\bibfield  {title}
  {\bibinfo {title} {Photoinduced $\ensuremath{\eta}$ pairing in the hubbard
  model},\ }\href {https://doi.org/10.1103/PhysRevLett.122.077002} {\bibfield
  {journal} {\bibinfo  {journal} {Phys. Rev. Lett.}\ }\textbf {\bibinfo
  {volume} {122}},\ \bibinfo {pages} {077002} (\bibinfo {year}
  {2019})}\BibitemShut {NoStop}%
\bibitem [{\citenamefont {Li}\ \emph {et~al.}(2020)\citenamefont {Li},
  \citenamefont {Golez}, \citenamefont {Werner},\ and\ \citenamefont
  {Eckstein}}]{Li2020}%
  \BibitemOpen
  \bibfield  {author} {\bibinfo {author} {\bibfnamefont {J.}~\bibnamefont
  {Li}}, \bibinfo {author} {\bibfnamefont {D.}~\bibnamefont {Golez}}, \bibinfo
  {author} {\bibfnamefont {P.}~\bibnamefont {Werner}},\ and\ \bibinfo {author}
  {\bibfnamefont {M.}~\bibnamefont {Eckstein}},\ }\bibfield  {title} {\bibinfo
  {title} {$\ensuremath{\eta}$-paired superconducting hidden phase in
  photodoped mott insulators},\ }\href
  {https://doi.org/10.1103/PhysRevB.102.165136} {\bibfield  {journal} {\bibinfo
   {journal} {Phys. Rev. B}\ }\textbf {\bibinfo {volume} {102}},\ \bibinfo
  {pages} {165136} (\bibinfo {year} {2020})}\BibitemShut {NoStop}%
\bibitem [{\citenamefont {Werner}\ and\ \citenamefont
  {Murakami}(2021)}]{Werner2021}%
  \BibitemOpen
  \bibfield  {author} {\bibinfo {author} {\bibfnamefont {P.}~\bibnamefont
  {Werner}}\ and\ \bibinfo {author} {\bibfnamefont {Y.}~\bibnamefont
  {Murakami}},\ }\bibfield  {title} {\bibinfo {title} {Light-induced hidden
  odd-frequency order in a model for ${A}_{3}{\mathrm{c}}_{60}$},\ }\href
  {https://doi.org/10.1103/PhysRevB.104.L201101} {\bibfield  {journal}
  {\bibinfo  {journal} {Phys. Rev. B}\ }\textbf {\bibinfo {volume} {104}},\
  \bibinfo {pages} {L201101} (\bibinfo {year} {2021})}\BibitemShut {NoStop}%
\bibitem [{\citenamefont {Ray}\ \emph {et~al.}(2023)\citenamefont {Ray},
  \citenamefont {Murakami},\ and\ \citenamefont {Werner}}]{Ray2023}%
  \BibitemOpen
  \bibfield  {author} {\bibinfo {author} {\bibfnamefont {S.}~\bibnamefont
  {Ray}}, \bibinfo {author} {\bibfnamefont {Y.}~\bibnamefont {Murakami}},\ and\
  \bibinfo {author} {\bibfnamefont {P.}~\bibnamefont {Werner}},\ }\bibfield
  {title} {\bibinfo {title} {Nonthermal superconductivity in photodoped
  multiorbital hubbard systems},\ }\href
  {https://doi.org/10.1103/PhysRevB.108.174515} {\bibfield  {journal} {\bibinfo
   {journal} {Phys. Rev. B}\ }\textbf {\bibinfo {volume} {108}},\ \bibinfo
  {pages} {174515} (\bibinfo {year} {2023})}\BibitemShut {NoStop}%
\bibitem [{\citenamefont {Ray}\ and\ \citenamefont {Werner}(2024)}]{Ray2024}%
  \BibitemOpen
  \bibfield  {author} {\bibinfo {author} {\bibfnamefont {S.}~\bibnamefont
  {Ray}}\ and\ \bibinfo {author} {\bibfnamefont {P.}~\bibnamefont {Werner}},\
  }\bibfield  {title} {\bibinfo {title} {Photoinduced charge, spin, and orbital
  order in the two-orbital extended hubbard model},\ }\href
  {https://doi.org/10.1103/PhysRevB.110.214433} {\bibfield  {journal} {\bibinfo
   {journal} {Phys. Rev. B}\ }\textbf {\bibinfo {volume} {110}},\ \bibinfo
  {pages} {214433} (\bibinfo {year} {2024})}\BibitemShut {NoStop}%
\bibitem [{\citenamefont {Freericks}\ \emph {et~al.}(2006)\citenamefont
  {Freericks}, \citenamefont {Turkowski},\ and\ \citenamefont
  {Zlati\ifmmode~\acute{c}\else \'{c}\fi{}}}]{Freericks2006}%
  \BibitemOpen
  \bibfield  {author} {\bibinfo {author} {\bibfnamefont {J.~K.}\ \bibnamefont
  {Freericks}}, \bibinfo {author} {\bibfnamefont {V.~M.}\ \bibnamefont
  {Turkowski}},\ and\ \bibinfo {author} {\bibfnamefont {V.}~\bibnamefont
  {Zlati\ifmmode~\acute{c}\else \'{c}\fi{}}},\ }\bibfield  {title} {\bibinfo
  {title} {Nonequilibrium dynamical mean-field theory},\ }\href
  {https://doi.org/10.1103/PhysRevLett.97.266408} {\bibfield  {journal}
  {\bibinfo  {journal} {Phys. Rev. Lett.}\ }\textbf {\bibinfo {volume} {97}},\
  \bibinfo {pages} {266408} (\bibinfo {year} {2006})}\BibitemShut {NoStop}%
\bibitem [{\citenamefont {Aoki}\ \emph {et~al.}(2014)\citenamefont {Aoki},
  \citenamefont {Tsuji}, \citenamefont {Eckstein}, \citenamefont {Kollar},
  \citenamefont {Oka},\ and\ \citenamefont {Werner}}]{Aoki2014}%
  \BibitemOpen
  \bibfield  {author} {\bibinfo {author} {\bibfnamefont {H.}~\bibnamefont
  {Aoki}}, \bibinfo {author} {\bibfnamefont {N.}~\bibnamefont {Tsuji}},
  \bibinfo {author} {\bibfnamefont {M.}~\bibnamefont {Eckstein}}, \bibinfo
  {author} {\bibfnamefont {M.}~\bibnamefont {Kollar}}, \bibinfo {author}
  {\bibfnamefont {T.}~\bibnamefont {Oka}},\ and\ \bibinfo {author}
  {\bibfnamefont {P.}~\bibnamefont {Werner}},\ }\bibfield  {title} {\bibinfo
  {title} {Nonequilibrium dynamical mean-field theory and its applications},\
  }\href {https://doi.org/10.1103/RevModPhys.86.779} {\bibfield  {journal}
  {\bibinfo  {journal} {Rev. Mod. Phys.}\ }\textbf {\bibinfo {volume} {86}},\
  \bibinfo {pages} {779} (\bibinfo {year} {2014})},\ \Eprint
  {https://arxiv.org/abs/1310.5329} {arXiv:1310.5329 [cond-mat.str-el]}
  \BibitemShut {NoStop}%
\bibitem [{\citenamefont {Bauer}\ \emph {et~al.}(2015)\citenamefont {Bauer},
  \citenamefont {Babadi},\ and\ \citenamefont {Demler}}]{Bauer2015}%
  \BibitemOpen
  \bibfield  {author} {\bibinfo {author} {\bibfnamefont {J.}~\bibnamefont
  {Bauer}}, \bibinfo {author} {\bibfnamefont {M.}~\bibnamefont {Babadi}},\ and\
  \bibinfo {author} {\bibfnamefont {E.}~\bibnamefont {Demler}},\ }\bibfield
  {title} {\bibinfo {title} {Dynamical instabilities and transient short-range
  order in the fermionic {Hubbard} model},\ }\href
  {https://doi.org/10.1103/PhysRevB.92.024305} {\bibfield  {journal} {\bibinfo
  {journal} {Phys. Rev. B}\ }\textbf {\bibinfo {volume} {92}},\ \bibinfo
  {pages} {024305} (\bibinfo {year} {2015})},\ \Eprint
  {https://arxiv.org/abs/1411.4049} {arXiv:1411.4049 [cond-mat.str-el]}
  \BibitemShut {NoStop}%
\bibitem [{\citenamefont {Bickers}\ \emph {et~al.}(1989)\citenamefont
  {Bickers}, \citenamefont {Scalapino},\ and\ \citenamefont
  {White}}]{Bickers1989}%
  \BibitemOpen
  \bibfield  {author} {\bibinfo {author} {\bibfnamefont {N.~E.}\ \bibnamefont
  {Bickers}}, \bibinfo {author} {\bibfnamefont {D.~J.}\ \bibnamefont
  {Scalapino}},\ and\ \bibinfo {author} {\bibfnamefont {S.~R.}\ \bibnamefont
  {White}},\ }\bibfield  {title} {\bibinfo {title} {Conserving approximations
  for strongly correlated electron systems: Bethe-salpeter equation and
  dynamics for the two-dimensional hubbard model},\ }\href
  {https://doi.org/10.1103/PhysRevLett.62.961} {\bibfield  {journal} {\bibinfo
  {journal} {Phys. Rev. Lett.}\ }\textbf {\bibinfo {volume} {62}},\ \bibinfo
  {pages} {961} (\bibinfo {year} {1989})}\BibitemShut {NoStop}%
\bibitem [{\citenamefont {Hedin}(1965)}]{Hedin1965}%
  \BibitemOpen
  \bibfield  {author} {\bibinfo {author} {\bibfnamefont {L.}~\bibnamefont
  {Hedin}},\ }\bibfield  {title} {\bibinfo {title} {New method for calculating
  the one-particle green's function with application to the electron-gas
  problem},\ }\href {https://doi.org/10.1103/PhysRev.139.A796} {\bibfield
  {journal} {\bibinfo  {journal} {Phys. Rev.}\ }\textbf {\bibinfo {volume}
  {139}},\ \bibinfo {pages} {A796} (\bibinfo {year} {1965})}\BibitemShut
  {NoStop}%
\bibitem [{\citenamefont {Migdal}(1958)}]{Migdal1958}%
  \BibitemOpen
  \bibfield  {author} {\bibinfo {author} {\bibfnamefont {A.~B.}\ \bibnamefont
  {Migdal}},\ }\bibfield  {title} {\bibinfo {title} {Interaction between
  electrons and lattice vibrations in a normal metal},\ }\href@noop {}
  {\bibfield  {journal} {\bibinfo  {journal} {Sov. Phys. JETP}\ }\textbf
  {\bibinfo {volume} {7}},\ \bibinfo {pages} {996} (\bibinfo {year}
  {1958})}\BibitemShut {NoStop}%
\bibitem [{\citenamefont {Sch{\"u}ler}\ \emph {et~al.}(2020)\citenamefont
  {Sch{\"u}ler}, \citenamefont {Gole{\v{z}}}, \citenamefont {Murakami},
  \citenamefont {Bittner}, \citenamefont {Herrmann}, \citenamefont {Strand},
  \citenamefont {Werner},\ and\ \citenamefont {Eckstein}}]{Schuler2019}%
  \BibitemOpen
  \bibfield  {author} {\bibinfo {author} {\bibfnamefont {M.}~\bibnamefont
  {Sch{\"u}ler}}, \bibinfo {author} {\bibfnamefont {D.}~\bibnamefont
  {Gole{\v{z}}}}, \bibinfo {author} {\bibfnamefont {Y.}~\bibnamefont
  {Murakami}}, \bibinfo {author} {\bibfnamefont {N.}~\bibnamefont {Bittner}},
  \bibinfo {author} {\bibfnamefont {A.}~\bibnamefont {Herrmann}}, \bibinfo
  {author} {\bibfnamefont {H.~U.~R.}\ \bibnamefont {Strand}}, \bibinfo {author}
  {\bibfnamefont {P.}~\bibnamefont {Werner}},\ and\ \bibinfo {author}
  {\bibfnamefont {M.}~\bibnamefont {Eckstein}},\ }\bibfield  {title} {\bibinfo
  {title} {{NESSi}: The non-equilibrium systems simulation package},\ }\href
  {https://doi.org/10.1016/j.cpc.2020.107484} {\bibfield  {journal} {\bibinfo
  {journal} {Comput. Phys. Commun.}\ }\textbf {\bibinfo {volume} {257}},\
  \bibinfo {pages} {107484} (\bibinfo {year} {2020})},\ \Eprint
  {https://arxiv.org/abs/1911.01211} {arXiv:1911.01211 [cs.CE]} \BibitemShut
  {NoStop}%
\bibitem [{\citenamefont {Shinaoka}\ \emph {et~al.}(2023)\citenamefont
  {Shinaoka}, \citenamefont {Wallerberger}, \citenamefont {Murakami},
  \citenamefont {Nogaki}, \citenamefont {Sakurai}, \citenamefont {Werner},\
  and\ \citenamefont {Kauch}}]{Shinaoka2023}%
  \BibitemOpen
  \bibfield  {author} {\bibinfo {author} {\bibfnamefont {H.}~\bibnamefont
  {Shinaoka}}, \bibinfo {author} {\bibfnamefont {M.}~\bibnamefont
  {Wallerberger}}, \bibinfo {author} {\bibfnamefont {Y.}~\bibnamefont
  {Murakami}}, \bibinfo {author} {\bibfnamefont {K.}~\bibnamefont {Nogaki}},
  \bibinfo {author} {\bibfnamefont {R.}~\bibnamefont {Sakurai}}, \bibinfo
  {author} {\bibfnamefont {P.}~\bibnamefont {Werner}},\ and\ \bibinfo {author}
  {\bibfnamefont {A.}~\bibnamefont {Kauch}},\ }\bibfield  {title} {\bibinfo
  {title} {Multiscale space-time ansatz for correlation functions of quantum
  systems based on quantics tensor trains},\ }\href
  {https://doi.org/10.1103/PhysRevX.13.021015} {\bibfield  {journal} {\bibinfo
  {journal} {Phys. Rev. X}\ }\textbf {\bibinfo {volume} {13}},\ \bibinfo
  {pages} {021015} (\bibinfo {year} {2023})},\ \Eprint
  {https://arxiv.org/abs/2210.12984} {arXiv:2210.12984 [cond-mat.str-el]}
  \BibitemShut {NoStop}%
\bibitem [{\citenamefont {Murray}\ \emph {et~al.}(2024)\citenamefont {Murray},
  \citenamefont {Shinaoka},\ and\ \citenamefont {Werner}}]{Murray2024}%
  \BibitemOpen
  \bibfield  {author} {\bibinfo {author} {\bibfnamefont {M.}~\bibnamefont
  {Murray}}, \bibinfo {author} {\bibfnamefont {H.}~\bibnamefont {Shinaoka}},\
  and\ \bibinfo {author} {\bibfnamefont {P.}~\bibnamefont {Werner}},\
  }\bibfield  {title} {\bibinfo {title} {Nonequilibrium diagrammatic many-body
  simulations with quantics tensor trains},\ }\href
  {https://doi.org/10.1103/PhysRevB.109.165135} {\bibfield  {journal} {\bibinfo
   {journal} {Phys. Rev. B}\ }\textbf {\bibinfo {volume} {109}},\ \bibinfo
  {pages} {165135} (\bibinfo {year} {2024})},\ \Eprint
  {https://arxiv.org/abs/2312.03809} {arXiv:2312.03809 [cond-mat.str-el]}
  \BibitemShut {NoStop}%
\bibitem [{\citenamefont {Hubbard}(1963)}]{Hubbard1963}%
  \BibitemOpen
  \bibfield  {author} {\bibinfo {author} {\bibfnamefont {J.}~\bibnamefont
  {Hubbard}},\ }\bibfield  {title} {\bibinfo {title} {Electron correlations in
  narrow energy bands},\ }\href {https://doi.org/10.1098/rspa.1963.0204}
  {\bibfield  {journal} {\bibinfo  {journal} {Proc. R. Soc. Lond. A}\ }\textbf
  {\bibinfo {volume} {276}},\ \bibinfo {pages} {238} (\bibinfo {year}
  {1963})}\BibitemShut {NoStop}%
\bibitem [{\citenamefont {Fr{\"o}hlich}(1950)}]{Frohlich1950}%
  \BibitemOpen
  \bibfield  {author} {\bibinfo {author} {\bibfnamefont {H.}~\bibnamefont
  {Fr{\"o}hlich}},\ }\bibfield  {title} {\bibinfo {title} {Theory of the
  superconducting state. {I}. {The} ground state at the absolute zero of
  temperature},\ }\href {https://doi.org/10.1103/PhysRev.79.845} {\bibfield
  {journal} {\bibinfo  {journal} {Phys. Rev.}\ }\textbf {\bibinfo {volume}
  {79}},\ \bibinfo {pages} {845} (\bibinfo {year} {1950})}\BibitemShut
  {NoStop}%
\bibitem [{\citenamefont {Bardeen}\ and\ \citenamefont
  {Shockley}(1950)}]{BardeenShockley1950}%
  \BibitemOpen
  \bibfield  {author} {\bibinfo {author} {\bibfnamefont {J.}~\bibnamefont
  {Bardeen}}\ and\ \bibinfo {author} {\bibfnamefont {W.}~\bibnamefont
  {Shockley}},\ }\bibfield  {title} {\bibinfo {title} {Deformation potentials
  and mobilities in non-polar crystals},\ }\href
  {https://doi.org/10.1103/PhysRev.80.72} {\bibfield  {journal} {\bibinfo
  {journal} {Phys. Rev.}\ }\textbf {\bibinfo {volume} {80}},\ \bibinfo {pages}
  {72} (\bibinfo {year} {1950})}\BibitemShut {NoStop}%
\bibitem [{\citenamefont {Schl{\"u}nzen}\ \emph {et~al.}(2020)\citenamefont
  {Schl{\"u}nzen}, \citenamefont {Hermanns}, \citenamefont {Scharnke},\ and\
  \citenamefont {Bonitz}}]{Schlunzen2020}%
  \BibitemOpen
  \bibfield  {author} {\bibinfo {author} {\bibfnamefont {N.}~\bibnamefont
  {Schl{\"u}nzen}}, \bibinfo {author} {\bibfnamefont {S.}~\bibnamefont
  {Hermanns}}, \bibinfo {author} {\bibfnamefont {M.}~\bibnamefont {Scharnke}},\
  and\ \bibinfo {author} {\bibfnamefont {M.}~\bibnamefont {Bonitz}},\
  }\bibfield  {title} {\bibinfo {title} {Ultrafast dynamics of strongly
  correlated fermions---nonequilibrium {Green} functions and selfenergy
  approximations},\ }\href {https://doi.org/10.1088/1361-648X/ab2d32}
  {\bibfield  {journal} {\bibinfo  {journal} {J. Phys.: Condens. Matter}\
  }\textbf {\bibinfo {volume} {32}},\ \bibinfo {pages} {103001} (\bibinfo
  {year} {2020})}\BibitemShut {NoStop}%
\bibitem [{\citenamefont {Stefanucci}\ and\ \citenamefont {van
  Leeuwen}(2025)}]{Stefanucci2025}%
  \BibitemOpen
  \bibfield  {author} {\bibinfo {author} {\bibfnamefont {G.}~\bibnamefont
  {Stefanucci}}\ and\ \bibinfo {author} {\bibfnamefont {R.}~\bibnamefont {van
  Leeuwen}},\ }\href {https://doi.org/10.1017/9781009536776} {\emph {\bibinfo
  {title} {Nonequilibrium Many-Body Theory of Quantum Systems: A Modern
  Introduction}}},\ \bibinfo {edition} {2nd}\ ed.\ (\bibinfo  {publisher}
  {Cambridge University Press},\ \bibinfo {address} {Cambridge},\ \bibinfo
  {year} {2025})\BibitemShut {NoStop}%
\bibitem [{\citenamefont {Bickers}\ and\ \citenamefont
  {Scalapino}(1989)}]{Bickers1989b}%
  \BibitemOpen
  \bibfield  {author} {\bibinfo {author} {\bibfnamefont {N.~E.}\ \bibnamefont
  {Bickers}}\ and\ \bibinfo {author} {\bibfnamefont {D.~J.}\ \bibnamefont
  {Scalapino}},\ }\bibfield  {title} {\bibinfo {title} {Conserving
  approximations for strongly fluctuating electron systems. {I}. formalism and
  calculational approach},\ }\href
  {https://doi.org/10.1016/0003-4916(89)90359-X} {\bibfield  {journal}
  {\bibinfo  {journal} {Ann. Phys.}\ }\textbf {\bibinfo {volume} {193}},\
  \bibinfo {pages} {206} (\bibinfo {year} {1989})}\BibitemShut {NoStop}%
\bibitem [{\citenamefont {Sayyad}\ \emph {et~al.}(2019)\citenamefont {Sayyad},
  \citenamefont {Tsuji}, \citenamefont {Vaezi}, \citenamefont {Capone},
  \citenamefont {Eckstein},\ and\ \citenamefont {Aoki}}]{Sayyad2019}%
  \BibitemOpen
  \bibfield  {author} {\bibinfo {author} {\bibfnamefont {S.}~\bibnamefont
  {Sayyad}}, \bibinfo {author} {\bibfnamefont {N.}~\bibnamefont {Tsuji}},
  \bibinfo {author} {\bibfnamefont {A.}~\bibnamefont {Vaezi}}, \bibinfo
  {author} {\bibfnamefont {M.}~\bibnamefont {Capone}}, \bibinfo {author}
  {\bibfnamefont {M.}~\bibnamefont {Eckstein}},\ and\ \bibinfo {author}
  {\bibfnamefont {H.}~\bibnamefont {Aoki}},\ }\bibfield  {title} {\bibinfo
  {title} {Momentum-dependent relaxation dynamics of the doped repulsive
  {Hubbard} model},\ }\href {https://doi.org/10.1103/PhysRevB.99.165132}
  {\bibfield  {journal} {\bibinfo  {journal} {Phys. Rev. B}\ }\textbf {\bibinfo
  {volume} {99}},\ \bibinfo {pages} {165132} (\bibinfo {year}
  {2019})}\BibitemShut {NoStop}%
\bibitem [{\citenamefont {Dasari}\ and\ \citenamefont
  {Eckstein}(2019)}]{Dasari2019}%
  \BibitemOpen
  \bibfield  {author} {\bibinfo {author} {\bibfnamefont {N.}~\bibnamefont
  {Dasari}}\ and\ \bibinfo {author} {\bibfnamefont {M.}~\bibnamefont
  {Eckstein}},\ }\bibfield  {title} {\bibinfo {title} {Ultrafast electric field
  controlled spin correlations in the {Hubbard} model},\ }\href
  {https://doi.org/10.1103/PhysRevB.100.121114} {\bibfield  {journal} {\bibinfo
   {journal} {Phys. Rev. B}\ }\textbf {\bibinfo {volume} {100}},\ \bibinfo
  {pages} {121114(R)} (\bibinfo {year} {2019})}\BibitemShut {NoStop}%
\bibitem [{\citenamefont {Monthoux}\ and\ \citenamefont
  {Scalapino}(1994)}]{Monthoux1994}%
  \BibitemOpen
  \bibfield  {author} {\bibinfo {author} {\bibfnamefont {P.}~\bibnamefont
  {Monthoux}}\ and\ \bibinfo {author} {\bibfnamefont {D.~J.}\ \bibnamefont
  {Scalapino}},\ }\bibfield  {title} {\bibinfo {title} {Self-consistent
  $d_{x^2-y^2}$ pairing in a two-dimensional {Hubbard} model},\ }\href
  {https://doi.org/10.1103/PhysRevLett.72.1874} {\bibfield  {journal} {\bibinfo
   {journal} {Phys. Rev. Lett.}\ }\textbf {\bibinfo {volume} {72}},\ \bibinfo
  {pages} {1874} (\bibinfo {year} {1994})}\BibitemShut {NoStop}%
\bibitem [{\citenamefont {Schmalian}\ \emph {et~al.}(1996)\citenamefont
  {Schmalian}, \citenamefont {Langer}, \citenamefont {Grabowski},\ and\
  \citenamefont {Bennemann}}]{Schmalian1996}%
  \BibitemOpen
  \bibfield  {author} {\bibinfo {author} {\bibfnamefont {J.}~\bibnamefont
  {Schmalian}}, \bibinfo {author} {\bibfnamefont {M.}~\bibnamefont {Langer}},
  \bibinfo {author} {\bibfnamefont {S.}~\bibnamefont {Grabowski}},\ and\
  \bibinfo {author} {\bibfnamefont {K.~H.}\ \bibnamefont {Bennemann}},\
  }\bibfield  {title} {\bibinfo {title} {Self-consistent summation of
  many-particle diagrams on the real frequency axis and its application to the
  {FLEX} approximation},\ }\href {https://doi.org/10.1016/0010-4655(95)00134-4}
  {\bibfield  {journal} {\bibinfo  {journal} {Comput. Phys. Commun.}\ }\textbf
  {\bibinfo {volume} {93}},\ \bibinfo {pages} {141} (\bibinfo {year}
  {1996})}\BibitemShut {NoStop}%
\bibitem [{\citenamefont {Geng}\ \emph {et~al.}(2025)\citenamefont {Geng},
  \citenamefont {Yan},\ and\ \citenamefont {Werner}}]{Geng2025}%
  \BibitemOpen
  \bibfield  {author} {\bibinfo {author} {\bibfnamefont {L.}~\bibnamefont
  {Geng}}, \bibinfo {author} {\bibfnamefont {J.}~\bibnamefont {Yan}},\ and\
  \bibinfo {author} {\bibfnamefont {P.}~\bibnamefont {Werner}},\ }\bibfield
  {title} {\bibinfo {title} {Two-particle self-consistent approach combined
  with dynamical mean field theory: A real-frequency study of the
  square-lattice {Hubbard} model},\ }\href
  {https://doi.org/10.1103/PhysRevB.111.115143} {\bibfield  {journal} {\bibinfo
   {journal} {Phys. Rev. B}\ }\textbf {\bibinfo {volume} {111}},\ \bibinfo
  {pages} {115143} (\bibinfo {year} {2025})}\BibitemShut {NoStop}%
\bibitem [{\citenamefont {Kitatani}\ \emph {et~al.}(2015)\citenamefont
  {Kitatani}, \citenamefont {Tsuji},\ and\ \citenamefont
  {Aoki}}]{Kitatani2015}%
  \BibitemOpen
  \bibfield  {author} {\bibinfo {author} {\bibfnamefont {M.}~\bibnamefont
  {Kitatani}}, \bibinfo {author} {\bibfnamefont {N.}~\bibnamefont {Tsuji}},\
  and\ \bibinfo {author} {\bibfnamefont {H.}~\bibnamefont {Aoki}},\ }\bibfield
  {title} {\bibinfo {title} {{FLEX+DMFT} approach to the $d$-wave
  superconducting phase diagram of the two-dimensional {Hubbard} model},\
  }\href {https://doi.org/10.1103/PhysRevB.92.085104} {\bibfield  {journal}
  {\bibinfo  {journal} {Phys. Rev. B}\ }\textbf {\bibinfo {volume} {92}},\
  \bibinfo {pages} {085104} (\bibinfo {year} {2015})}\BibitemShut {NoStop}%
\bibitem [{\citenamefont {Mermin}\ and\ \citenamefont
  {Wagner}(1966)}]{Mermin1966}%
  \BibitemOpen
  \bibfield  {author} {\bibinfo {author} {\bibfnamefont {N.~D.}\ \bibnamefont
  {Mermin}}\ and\ \bibinfo {author} {\bibfnamefont {H.}~\bibnamefont
  {Wagner}},\ }\bibfield  {title} {\bibinfo {title} {Absence of ferromagnetism
  or antiferromagnetism in one- or two-dimensional isotropic {H}eisenberg
  models},\ }\href {https://doi.org/10.1103/PhysRevLett.17.1133} {\bibfield
  {journal} {\bibinfo  {journal} {Phys. Rev. Lett.}\ }\textbf {\bibinfo
  {volume} {17}},\ \bibinfo {pages} {1133} (\bibinfo {year}
  {1966})}\BibitemShut {NoStop}%
\bibitem [{\citenamefont {Ghosh}(1971)}]{Ghosh1971}%
  \BibitemOpen
  \bibfield  {author} {\bibinfo {author} {\bibfnamefont {D.~K.}\ \bibnamefont
  {Ghosh}},\ }\bibfield  {title} {\bibinfo {title} {Nonexistence of magnetic
  ordering in the one- and two-dimensional {H}ubbard model},\ }\href
  {https://doi.org/10.1103/PhysRevLett.27.1584} {\bibfield  {journal} {\bibinfo
   {journal} {Phys. Rev. Lett.}\ }\textbf {\bibinfo {volume} {27}},\ \bibinfo
  {pages} {1584} (\bibinfo {year} {1971})}\BibitemShut {NoStop}%
\bibitem [{\citenamefont {Kontani}\ and\ \citenamefont
  {Ohno}(2006)}]{Kontani2006}%
  \BibitemOpen
  \bibfield  {author} {\bibinfo {author} {\bibfnamefont {H.}~\bibnamefont
  {Kontani}}\ and\ \bibinfo {author} {\bibfnamefont {M.}~\bibnamefont {Ohno}},\
  }\bibfield  {title} {\bibinfo {title} {Effect of a nonmagnetic impurity in a
  nearly antiferromagnetic {F}ermi liquid: Magnetic correlations and transport
  phenomena},\ }\href {https://doi.org/10.1103/PhysRevB.74.014406} {\bibfield
  {journal} {\bibinfo  {journal} {Phys. Rev. B}\ }\textbf {\bibinfo {volume}
  {74}},\ \bibinfo {pages} {014406} (\bibinfo {year} {2006})},\ \Eprint
  {https://arxiv.org/abs/cond-mat/0607595} {arXiv:cond-mat/0607595}
  \BibitemShut {NoStop}%
\bibitem [{\citenamefont {Deisz}\ \emph {et~al.}(1996)\citenamefont {Deisz},
  \citenamefont {Hess},\ and\ \citenamefont {Serene}}]{Deisz1996}%
  \BibitemOpen
  \bibfield  {author} {\bibinfo {author} {\bibfnamefont {J.~J.}\ \bibnamefont
  {Deisz}}, \bibinfo {author} {\bibfnamefont {D.~W.}\ \bibnamefont {Hess}},\
  and\ \bibinfo {author} {\bibfnamefont {J.~W.}\ \bibnamefont {Serene}},\
  }\bibfield  {title} {\bibinfo {title} {Incipient antiferromagnetism and
  low-energy excitations in the half-filled two-dimensional {H}ubbard model},\
  }\href {https://doi.org/10.1103/PhysRevLett.76.1312} {\bibfield  {journal}
  {\bibinfo  {journal} {Phys. Rev. Lett.}\ }\textbf {\bibinfo {volume} {76}},\
  \bibinfo {pages} {1312} (\bibinfo {year} {1996})},\ \Eprint
  {https://arxiv.org/abs/cond-mat/9512084} {arXiv:cond-mat/9512084}
  \BibitemShut {NoStop}%
\bibitem [{\citenamefont {Takimoto}\ and\ \citenamefont
  {Moriya}(2002)}]{Takimoto2002}%
  \BibitemOpen
  \bibfield  {author} {\bibinfo {author} {\bibfnamefont {T.}~\bibnamefont
  {Takimoto}}\ and\ \bibinfo {author} {\bibfnamefont {T.}~\bibnamefont
  {Moriya}},\ }\bibfield  {title} {\bibinfo {title} {Superconductivity and
  antiferromagnetism in the three-dimensional {Hubbard} model},\ }\href
  {https://doi.org/10.1103/PhysRevB.66.134516} {\bibfield  {journal} {\bibinfo
  {journal} {Phys. Rev. B}\ }\textbf {\bibinfo {volume} {66}},\ \bibinfo
  {pages} {134516} (\bibinfo {year} {2002})}\BibitemShut {NoStop}%
\bibitem [{\citenamefont {Sakurazawa}\ \emph {et~al.}(2005)\citenamefont
  {Sakurazawa}, \citenamefont {Kontani},\ and\ \citenamefont
  {Saso}}]{Sakurazawa2005}%
  \BibitemOpen
  \bibfield  {author} {\bibinfo {author} {\bibfnamefont {K.}~\bibnamefont
  {Sakurazawa}}, \bibinfo {author} {\bibfnamefont {H.}~\bibnamefont
  {Kontani}},\ and\ \bibinfo {author} {\bibfnamefont {T.}~\bibnamefont
  {Saso}},\ }\bibfield  {title} {\bibinfo {title} {Magnetic-field-induced
  antiferromagnetism in two-dimensional {Hubbard} model: Analysis of
  {CeRhIn$_5$}},\ }\href {https://doi.org/10.1143/JPSJ.74.271} {\bibfield
  {journal} {\bibinfo  {journal} {J. Phys. Soc. Jpn.}\ }\textbf {\bibinfo
  {volume} {74}},\ \bibinfo {pages} {271} (\bibinfo {year} {2005})}\BibitemShut
  {NoStop}%
\bibitem [{\citenamefont {Hatakeyama}\ and\ \citenamefont
  {Ikeda}(2014)}]{Hatakeyama2014}%
  \BibitemOpen
  \bibfield  {author} {\bibinfo {author} {\bibfnamefont {Y.}~\bibnamefont
  {Hatakeyama}}\ and\ \bibinfo {author} {\bibfnamefont {R.}~\bibnamefont
  {Ikeda}},\ }\bibfield  {title} {\bibinfo {title} {Strong-coupling approach to
  antiferromagnetic ordering driven by paramagnetic pair-breaking in $d$-wave
  superconducting phase},\ }\href {https://doi.org/10.7566/JPSJ.83.024713}
  {\bibfield  {journal} {\bibinfo  {journal} {J. Phys. Soc. Jpn.}\ }\textbf
  {\bibinfo {volume} {83}},\ \bibinfo {pages} {024713} (\bibinfo {year}
  {2014})}\BibitemShut {NoStop}%
\bibitem [{\citenamefont {Nakamura}\ and\ \citenamefont
  {Yanase}(2026)}]{Nakamura2026}%
  \BibitemOpen
  \bibfield  {author} {\bibinfo {author} {\bibfnamefont {K.}~\bibnamefont
  {Nakamura}}\ and\ \bibinfo {author} {\bibfnamefont {Y.}~\bibnamefont
  {Yanase}},\ }\bibfield  {title} {\bibinfo {title} {Supercurrent-induced
  antiferromagnetic order and spin-triplet pair generation in quantum critical
  $d$-wave superconductors},\ }\href {https://doi.org/10.1103/ghhk-2tm7}
  {\bibfield  {journal} {\bibinfo  {journal} {Phys. Rev. B}\ }\textbf {\bibinfo
  {volume} {113}},\ \bibinfo {pages} {014508} (\bibinfo {year}
  {2026})}\BibitemShut {NoStop}%
\bibitem [{\citenamefont {Oka}\ and\ \citenamefont {Aoki}(2010)}]{Oka2010}%
  \BibitemOpen
  \bibfield  {author} {\bibinfo {author} {\bibfnamefont {T.}~\bibnamefont
  {Oka}}\ and\ \bibinfo {author} {\bibfnamefont {H.}~\bibnamefont {Aoki}},\
  }\bibfield  {title} {\bibinfo {title} {Nonequilibrium superconducting and
  magnetic phases in a correlated electron system coupled to electrodes},\
  }\href {https://doi.org/10.1103/PhysRevB.82.064516} {\bibfield  {journal}
  {\bibinfo  {journal} {Phys. Rev. B}\ }\textbf {\bibinfo {volume} {82}},\
  \bibinfo {pages} {064516} (\bibinfo {year} {2010})}\BibitemShut {NoStop}%
\bibitem [{\citenamefont {Abdurazakov}\ \emph {et~al.}(2018)\citenamefont
  {Abdurazakov}, \citenamefont {Nevola}, \citenamefont {Rustagi}, \citenamefont
  {Freericks}, \citenamefont {Dougherty},\ and\ \citenamefont
  {Kemper}}]{Abdurazakov2018}%
  \BibitemOpen
  \bibfield  {author} {\bibinfo {author} {\bibfnamefont {O.}~\bibnamefont
  {Abdurazakov}}, \bibinfo {author} {\bibfnamefont {D.}~\bibnamefont {Nevola}},
  \bibinfo {author} {\bibfnamefont {A.}~\bibnamefont {Rustagi}}, \bibinfo
  {author} {\bibfnamefont {J.~K.}\ \bibnamefont {Freericks}}, \bibinfo {author}
  {\bibfnamefont {D.~B.}\ \bibnamefont {Dougherty}},\ and\ \bibinfo {author}
  {\bibfnamefont {A.~F.}\ \bibnamefont {Kemper}},\ }\bibfield  {title}
  {\bibinfo {title} {Nonequilibrium electron dynamics in pump-probe
  spectroscopy: Role of excited phonon populations},\ }\href
  {https://doi.org/10.1103/PhysRevB.98.245110} {\bibfield  {journal} {\bibinfo
  {journal} {Phys. Rev. B}\ }\textbf {\bibinfo {volume} {98}},\ \bibinfo
  {pages} {245110} (\bibinfo {year} {2018})}\BibitemShut {NoStop}%
\bibitem [{\citenamefont {Baym}(1962)}]{Baym1962}%
  \BibitemOpen
  \bibfield  {author} {\bibinfo {author} {\bibfnamefont {G.}~\bibnamefont
  {Baym}},\ }\bibfield  {title} {\bibinfo {title} {Self-consistent
  approximations in many-body systems},\ }\href
  {https://doi.org/10.1103/PhysRev.127.1391} {\bibfield  {journal} {\bibinfo
  {journal} {Phys. Rev.}\ }\textbf {\bibinfo {volume} {127}},\ \bibinfo {pages}
  {1391} (\bibinfo {year} {1962})}\BibitemShut {NoStop}%
\bibitem [{\citenamefont {Inayoshi}\ \emph {et~al.}(2026)\citenamefont
  {Inayoshi}, \citenamefont {{\'S}roda}, \citenamefont {Kauch}, \citenamefont
  {Werner},\ and\ \citenamefont {Shinaoka}}]{Inayoshi2025}%
  \BibitemOpen
  \bibfield  {author} {\bibinfo {author} {\bibfnamefont {K.}~\bibnamefont
  {Inayoshi}}, \bibinfo {author} {\bibfnamefont {M.}~\bibnamefont {{\'S}roda}},
  \bibinfo {author} {\bibfnamefont {A.}~\bibnamefont {Kauch}}, \bibinfo
  {author} {\bibfnamefont {P.}~\bibnamefont {Werner}},\ and\ \bibinfo {author}
  {\bibfnamefont {H.}~\bibnamefont {Shinaoka}},\ }\bibfield  {title} {\bibinfo
  {title} {A causality-based divide-and-conquer algorithm for nonequilibrium
  {G}reen's function calculations with quantics tensor trains},\ }\href
  {https://doi.org/10.21468/SciPostPhys.20.3.077} {\bibfield  {journal}
  {\bibinfo  {journal} {SciPost Phys.}\ }\textbf {\bibinfo {volume} {20}},\
  \bibinfo {pages} {077} (\bibinfo {year} {2026})},\ \Eprint
  {https://arxiv.org/abs/2509.15028} {arXiv:2509.15028 [cond-mat.str-el]}
  \BibitemShut {NoStop}%
\bibitem [{\citenamefont {{\'S}roda}\ \emph {et~al.}(2026)\citenamefont
  {{\'S}roda}, \citenamefont {Inayoshi}, \citenamefont {Sch{\"u}ler},
  \citenamefont {Shinaoka},\ and\ \citenamefont {Werner}}]{Sroda2025b}%
  \BibitemOpen
  \bibfield  {author} {\bibinfo {author} {\bibfnamefont {M.}~\bibnamefont
  {{\'S}roda}}, \bibinfo {author} {\bibfnamefont {K.}~\bibnamefont {Inayoshi}},
  \bibinfo {author} {\bibfnamefont {M.}~\bibnamefont {Sch{\"u}ler}}, \bibinfo
  {author} {\bibfnamefont {H.}~\bibnamefont {Shinaoka}},\ and\ \bibinfo
  {author} {\bibfnamefont {P.}~\bibnamefont {Werner}},\ }\bibfield  {title}
  {\bibinfo {title} {Predictor-corrector method based on dynamic mode
  decomposition for tensor-train nonequilibrium {G}reen's function
  calculations},\ }\href {https://doi.org/10.1103/xnf8-kntz} {\bibfield
  {journal} {\bibinfo  {journal} {Phys. Rev. B}\ }\textbf {\bibinfo {volume}
  {113}},\ \bibinfo {pages} {165113} (\bibinfo {year} {2026})},\ \Eprint
  {https://arxiv.org/abs/2509.22177} {arXiv:2509.22177 [cond-mat.str-el]}
  \BibitemShut {NoStop}%
\bibitem [{\citenamefont {Bezanson}\ \emph {et~al.}(2017)\citenamefont
  {Bezanson}, \citenamefont {Edelman}, \citenamefont {Karpinski},\ and\
  \citenamefont {Shah}}]{bezanson2017julia}%
  \BibitemOpen
  \bibfield  {author} {\bibinfo {author} {\bibfnamefont {J.}~\bibnamefont
  {Bezanson}}, \bibinfo {author} {\bibfnamefont {A.}~\bibnamefont {Edelman}},
  \bibinfo {author} {\bibfnamefont {S.}~\bibnamefont {Karpinski}},\ and\
  \bibinfo {author} {\bibfnamefont {V.~B.}\ \bibnamefont {Shah}},\ }\bibfield
  {title} {\bibinfo {title} {Julia: A fresh approach to numerical computing},\
  }\href {https://doi.org/10.1137/141000671} {\bibfield  {journal} {\bibinfo
  {journal} {SIAM Rev.}\ }\textbf {\bibinfo {volume} {59}},\ \bibinfo {pages}
  {65} (\bibinfo {year} {2017})}\BibitemShut {NoStop}%
\bibitem [{\citenamefont {Fishman}\ \emph {et~al.}(2022)\citenamefont
  {Fishman}, \citenamefont {White},\ and\ \citenamefont
  {Stoudenmire}}]{itensor}%
  \BibitemOpen
  \bibfield  {author} {\bibinfo {author} {\bibfnamefont {M.}~\bibnamefont
  {Fishman}}, \bibinfo {author} {\bibfnamefont {S.~R.}\ \bibnamefont {White}},\
  and\ \bibinfo {author} {\bibfnamefont {E.~M.}\ \bibnamefont {Stoudenmire}},\
  }\bibfield  {title} {\bibinfo {title} {{The ITensor Software Library for
  Tensor Network Calculations}},\ }\href
  {https://doi.org/10.21468/SciPostPhysCodeb.4} {\bibfield  {journal} {\bibinfo
   {journal} {SciPost Phys. Codebases}\ ,\ \bibinfo {pages} {4}} (\bibinfo
  {year} {2022})}\BibitemShut {NoStop}%
\bibitem [{\citenamefont {N{\'u}{\~n}ez~Fern{\'a}ndez}\ \emph
  {et~al.}(2025)\citenamefont {N{\'u}{\~n}ez~Fern{\'a}ndez}, \citenamefont
  {Ritter}, \citenamefont {Jeannin}, \citenamefont {Li}, \citenamefont {Kloss},
  \citenamefont {Louvet}, \citenamefont {Terasaki}, \citenamefont {Parcollet},
  \citenamefont {von Delft}, \citenamefont {Shinaoka},\ and\ \citenamefont
  {Waintal}}]{Fernandez2024}%
  \BibitemOpen
  \bibfield  {author} {\bibinfo {author} {\bibfnamefont {Y.}~\bibnamefont
  {N{\'u}{\~n}ez~Fern{\'a}ndez}}, \bibinfo {author} {\bibfnamefont {M.~K.}\
  \bibnamefont {Ritter}}, \bibinfo {author} {\bibfnamefont {M.}~\bibnamefont
  {Jeannin}}, \bibinfo {author} {\bibfnamefont {J.-W.}\ \bibnamefont {Li}},
  \bibinfo {author} {\bibfnamefont {T.}~\bibnamefont {Kloss}}, \bibinfo
  {author} {\bibfnamefont {T.}~\bibnamefont {Louvet}}, \bibinfo {author}
  {\bibfnamefont {S.}~\bibnamefont {Terasaki}}, \bibinfo {author}
  {\bibfnamefont {O.}~\bibnamefont {Parcollet}}, \bibinfo {author}
  {\bibfnamefont {J.}~\bibnamefont {von Delft}}, \bibinfo {author}
  {\bibfnamefont {H.}~\bibnamefont {Shinaoka}},\ and\ \bibinfo {author}
  {\bibfnamefont {X.}~\bibnamefont {Waintal}},\ }\bibfield  {title} {\bibinfo
  {title} {{Learning tensor networks with tensor cross interpolation: New
  algorithms and libraries}},\ }\href
  {https://doi.org/10.21468/SciPostPhys.18.3.104} {\bibfield  {journal}
  {\bibinfo  {journal} {SciPost Phys.}\ }\textbf {\bibinfo {volume} {18}},\
  \bibinfo {pages} {104} (\bibinfo {year} {2025})}\BibitemShut {NoStop}%
\bibitem [{ten()}]{tensor4all}%
  \BibitemOpen
  \href@noop {} {\bibinfo {title} {Tensor4all}},\ \bibinfo {howpublished}
  {\url{https://tensor4all.org}}\BibitemShut {NoStop}%
\bibitem [{\citenamefont {LeBlanc}\ \emph {et~al.}(2019)\citenamefont
  {LeBlanc}, \citenamefont {Li}, \citenamefont {Chen}, \citenamefont {Levy},
  \citenamefont {Antipov}, \citenamefont {Millis},\ and\ \citenamefont
  {Gull}}]{LeBlanc2019}%
  \BibitemOpen
  \bibfield  {author} {\bibinfo {author} {\bibfnamefont {J.~P.~F.}\
  \bibnamefont {LeBlanc}}, \bibinfo {author} {\bibfnamefont {S.}~\bibnamefont
  {Li}}, \bibinfo {author} {\bibfnamefont {X.}~\bibnamefont {Chen}}, \bibinfo
  {author} {\bibfnamefont {R.}~\bibnamefont {Levy}}, \bibinfo {author}
  {\bibfnamefont {A.~E.}\ \bibnamefont {Antipov}}, \bibinfo {author}
  {\bibfnamefont {A.~J.}\ \bibnamefont {Millis}},\ and\ \bibinfo {author}
  {\bibfnamefont {E.}~\bibnamefont {Gull}},\ }\bibfield  {title} {\bibinfo
  {title} {Magnetic susceptibility and simulated neutron signal in the
  two-dimensional {Hubbard} model},\ }\href
  {https://doi.org/10.1103/PhysRevB.100.075123} {\bibfield  {journal} {\bibinfo
   {journal} {Phys. Rev. B}\ }\textbf {\bibinfo {volume} {100}},\ \bibinfo
  {pages} {075123} (\bibinfo {year} {2019})}\BibitemShut {NoStop}%
\bibitem [{\citenamefont {Sch{\"a}fer}\ \emph {et~al.}(2021)\citenamefont
  {Sch{\"a}fer}, \citenamefont {Wentzell}, \citenamefont {{\v{S}}imkovic},
  \citenamefont {He}, \citenamefont {Hille}, \citenamefont {Klett},
  \citenamefont {Eckhardt}, \citenamefont {Arzhang}, \citenamefont {Harkov},
  \citenamefont {Le~R{\'e}gent}, \citenamefont {Kirsch}, \citenamefont {Wang},
  \citenamefont {Kim}, \citenamefont {Kozik}, \citenamefont {Stepanov},
  \citenamefont {Kauch}, \citenamefont {Andergassen}, \citenamefont {Hansmann},
  \citenamefont {Rohe}, \citenamefont {Vilk}, \citenamefont {LeBlanc},
  \citenamefont {Zhang}, \citenamefont {Tremblay}, \citenamefont {Ferrero},
  \citenamefont {Parcollet},\ and\ \citenamefont {Georges}}]{Schafer2021}%
  \BibitemOpen
  \bibfield  {author} {\bibinfo {author} {\bibfnamefont {T.}~\bibnamefont
  {Sch{\"a}fer}}, \bibinfo {author} {\bibfnamefont {N.}~\bibnamefont
  {Wentzell}}, \bibinfo {author} {\bibfnamefont {F.}~\bibnamefont
  {{\v{S}}imkovic}, \bibfnamefont {IV}}, \bibinfo {author} {\bibfnamefont
  {Y.-Y.}\ \bibnamefont {He}}, \bibinfo {author} {\bibfnamefont
  {C.}~\bibnamefont {Hille}}, \bibinfo {author} {\bibfnamefont
  {M.}~\bibnamefont {Klett}}, \bibinfo {author} {\bibfnamefont {C.~J.}\
  \bibnamefont {Eckhardt}}, \bibinfo {author} {\bibfnamefont {B.}~\bibnamefont
  {Arzhang}}, \bibinfo {author} {\bibfnamefont {V.}~\bibnamefont {Harkov}},
  \bibinfo {author} {\bibfnamefont {F.-M.}\ \bibnamefont {Le~R{\'e}gent}},
  \bibinfo {author} {\bibfnamefont {A.}~\bibnamefont {Kirsch}}, \bibinfo
  {author} {\bibfnamefont {Y.}~\bibnamefont {Wang}}, \bibinfo {author}
  {\bibfnamefont {A.~J.}\ \bibnamefont {Kim}}, \bibinfo {author} {\bibfnamefont
  {E.}~\bibnamefont {Kozik}}, \bibinfo {author} {\bibfnamefont {E.~A.}\
  \bibnamefont {Stepanov}}, \bibinfo {author} {\bibfnamefont {A.}~\bibnamefont
  {Kauch}}, \bibinfo {author} {\bibfnamefont {S.}~\bibnamefont {Andergassen}},
  \bibinfo {author} {\bibfnamefont {P.}~\bibnamefont {Hansmann}}, \bibinfo
  {author} {\bibfnamefont {D.}~\bibnamefont {Rohe}}, \bibinfo {author}
  {\bibfnamefont {Y.~M.}\ \bibnamefont {Vilk}}, \bibinfo {author}
  {\bibfnamefont {J.~P.~F.}\ \bibnamefont {LeBlanc}}, \bibinfo {author}
  {\bibfnamefont {S.}~\bibnamefont {Zhang}}, \bibinfo {author} {\bibfnamefont
  {A.-M.~S.}\ \bibnamefont {Tremblay}}, \bibinfo {author} {\bibfnamefont
  {M.}~\bibnamefont {Ferrero}}, \bibinfo {author} {\bibfnamefont
  {O.}~\bibnamefont {Parcollet}},\ and\ \bibinfo {author} {\bibfnamefont
  {A.}~\bibnamefont {Georges}},\ }\bibfield  {title} {\bibinfo {title}
  {Tracking the footprints of spin fluctuations: {A} {MultiMethod},
  {MultiMessenger} study of the two-dimensional {Hubbard} model},\ }\href
  {https://doi.org/10.1103/PhysRevX.11.011058} {\bibfield  {journal} {\bibinfo
  {journal} {Phys. Rev. X}\ }\textbf {\bibinfo {volume} {11}},\ \bibinfo
  {pages} {011058} (\bibinfo {year} {2021})},\ \Eprint
  {https://arxiv.org/abs/2006.10769} {arXiv:2006.10769 [cond-mat.str-el]}
  \BibitemShut {NoStop}%
\bibitem [{\citenamefont {Weber}\ and\ \citenamefont
  {Hohenadler}(2018)}]{Weber2018}%
  \BibitemOpen
  \bibfield  {author} {\bibinfo {author} {\bibfnamefont {M.}~\bibnamefont
  {Weber}}\ and\ \bibinfo {author} {\bibfnamefont {M.}~\bibnamefont
  {Hohenadler}},\ }\bibfield  {title} {\bibinfo {title} {Two-dimensional
  {Holstein-Hubbard} model: Critical temperature, {Ising} universality, and
  bipolaron liquid},\ }\href {https://doi.org/10.1103/PhysRevB.98.085405}
  {\bibfield  {journal} {\bibinfo  {journal} {Phys. Rev. B}\ }\textbf {\bibinfo
  {volume} {98}},\ \bibinfo {pages} {085405} (\bibinfo {year}
  {2018})}\BibitemShut {NoStop}%
\bibitem [{\citenamefont {Gukelberger}\ \emph {et~al.}(2015)\citenamefont
  {Gukelberger}, \citenamefont {Huang},\ and\ \citenamefont
  {Werner}}]{Gukelberger2015}%
  \BibitemOpen
  \bibfield  {author} {\bibinfo {author} {\bibfnamefont {J.}~\bibnamefont
  {Gukelberger}}, \bibinfo {author} {\bibfnamefont {L.}~\bibnamefont {Huang}},\
  and\ \bibinfo {author} {\bibfnamefont {P.}~\bibnamefont {Werner}},\
  }\bibfield  {title} {\bibinfo {title} {On the dangers of partial diagrammatic
  summations: Benchmarks for the two-dimensional {Hubbard} model in the
  weak-coupling regime},\ }\href {https://doi.org/10.1103/PhysRevB.91.235114}
  {\bibfield  {journal} {\bibinfo  {journal} {Phys. Rev. B}\ }\textbf {\bibinfo
  {volume} {91}},\ \bibinfo {pages} {235114} (\bibinfo {year}
  {2015})}\BibitemShut {NoStop}%
\end{thebibliography}%

\end{document}